\documentclass[reprint,aps,prb.superscriptaddress,amsmath,amssymb,twocolumn,showpacs]{revtex4-2}
\usepackage[dvipdfmx]{graphicx, hyperref, xcolor}
\usepackage{amsmath, amssymb, amsfonts, mathtools, amsthm, bm, here, comment, mathrsfs, pxrubrica, physics, subfigure, url, bbding}
\usepackage[version=4]{mhchem}
\begin{document}
\title{Systematic study for two-dimensional $Z_2$ topological phase transitions at high-symmetry points in all layer groups}
\author{Ren Sasaki}
\author{Yutaro Tanaka}
\author{Shuichi Murakami}
\affiliation{Department of Physics, Tokyo Institute of Technology, 2-12-1 Ookayama, Meguro-ku, Tokyo 152-8551, Japan}
\date{\today}

\begin{abstract}
    We construct a general theory of $Z_2$ topological phase transitions in two-dimensional systems with time-reversal symmetry.
    We investigate the possibilities of $Z_2$ topological phase transitions at band inversions at all high-symmetry points in $k$-space in all the 80 layer groups.
    We exclude the layer groups with inversion symmetry because the $Z_2$ topological phase transition is known to be associated with band inversions with an exchange of parities.
    Among the other layer groups, we find 21 layer groups with insulator-to-insulator transitions with band inversion, and this problem is finally reduced to five point groups $C_3, C_4, C_6, S_4$, and $C_{3h}$.
    We show how the change of the $Z_2$ topological invariant at a band inversion is entirely determined by the irreps of occupied and unoccupied bands at the high-symmetry point.
    For example, in the case of $C_3$, we show that the $Z_2$ topological invariants change whenever the band inversion occurs between two Kramers pairs whose $C_3$ eigenvalues are $\{e^{\pi i / 3}, e^{-\pi i / 3}\}$ and $\{-1, -1\}$.
    These results are not included in the theory of symmetry-based indicators or topological quantum chemistry.
\end{abstract}

\maketitle

\section{introduction}\label{sec:introduction}
Topology has become a key concept in our current understanding of condensed matter physics.
In particular, since the theory of two-dimensional (2D) topological insulators in 2005~\cite{Kane2005-ho}, various topological materials such as $Z_2$ topological insulators~\cite{Kane-Mele-Graphene,Bernevig2006-rq,Bernevig-2006,Murakami-2006,Fu-Kane-Mele,Konig2007-av,Moore2007-lr,Roy-3dti,Hasan2010-dk,Qi2011-hm,Fu2011-ar,Hsieh2008,Xia2009,Zhang2009}, topological superconductors~\cite{Read_2000,Kitaev_2001,Fu-2008-sc,Sasaki2011-xm,Sato2017-vl} and topological semimetals~\cite{Murakami2007-gu,Wan2011-og,Weng-2015,Su-weyl,Lv-weyl,Lv2015,Yang2015,Murakami2017-eh} have been proposed and verified both theoretically and experimentally.
These materials are characterized by the nontrivial topology of their electronic band structure.
It is essential to utilize the symmetry of the materials to elucidate the topology.
For example, in topological insulators protected by time-reversal symmetry, band inversion is induced by spin--orbit coupling that preserves time-reversal symmetry, resulting in a different topology of band structure from that of ordinary insulators.
The topology of the band structure is then characterized by $Z_2$ topological invariants expressed in terms of eigenstates of the systems~\cite{Fu2006-so}.
The expression is written as integrals in terms of the eigenstates over the entire Brillouin zone, which reflects global information in momentum space.
Meanwhile, the expression of topological invariants can be simplified in the presence of additional symmetries.
In particular, when the 2D system has inversion symmetry, the $Z_2$ topological invariants $\nu$ can be written in terms of parity eigenvalues at high-symmetry points, called time-reversal invariant momenta (TRIM), below the Fermi energy~\cite{FuKane-inversion}, which significantly simplifies the calculation.
On the other hand, if the system has no inversion symmetry, it is not simple to compute the $Z_2$ topological invariants.
In this case, it seems impossible to obtain topological information from only the information at high-symmetry points, unlike the case of inversion symmetry.
However, this paper shows that even in cases without inversion symmetry, a change of $Z_2$ topological invariant at band inversion between different irreducible representations (irreps) is directly related to the irreps involved.

This study aims to investigate whether the $Z_2$ topological phase transition, \textit{i}.\textit{e}., a phase transition from a normal insulator to a topological insulator, occurs at band inversions under various symmetries in two-dimensional spinful systems with time-reversal symmetry.
As mentioned above, in systems with inversion symmetry, the $Z_2$ topological invariant is written in terms of parity eigenvalues at TRIM. It is an example
of the symmetry-based indicator and topological quantum chemistry \cite{Po2017-mk,Kruthoff2017-zh,Bradlyn2017-ot,Bradlyn2018-eu,Khalaf-PRX}.
This simple formula tells us what kind of band-gap closings occur and how they are related to $Z_2$ topological phase transitions.
Namely, as we change a parameter in a system with inversion symmetry, band inversion between bands with opposite parities at a TRIM is always accompanied by a $Z_2$ topological phase transition. Meanwhile, the band gap never closes at non-TRIM points in inversion-symmetric systems because the co-dimension for the gap closing at non-TRIM points is equal to five in this case, and it is larger than the number of parameters ($k_x,k_y,m$) in the system, where $m$ is a parameter in the system \cite{Murakami2007-gu,Bzdusek}. On the other hand,
in two-dimensional systems without inversion symmetry, the $Z_2$ topological invariant cannot be written in terms of irreducible representations (irreps) at high-symmetry
momenta, and it is outside the theory of the symmetry-based indicator and topological quantum chemistry.
This is closely related to the fact that the $Z_2$ topological phase transition occurs by gap closing at general $\bm{k}$ points~\cite{Murakami2007-lt,Murakami2007-gu,Yu2020-ik}.

\begin{figure}[tbp]
    \centering
    \includegraphics[width = 1\columnwidth]{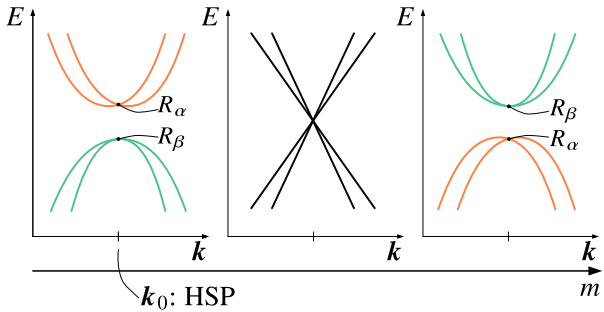}
    \caption{Schematic of an insulator-to-insulator transition. The three graphs in this figure illustrate the band structure of a 2D model for different values of a tunable parameter $m$. As we increase the value of $m$, the band gap closes, and the band inversion between two irreps $R_\alpha$ and $R_\beta$ occurs at an HSP $\bm{k}_0$ at a particular value of $m$. }
    \label{fig:schematic_of_phase_transition}
\end{figure}

In this paper, we focus on a system with an additional crystallographic symmetry (but not inversion symmetry); then, the band gap can close at high-symmetry points via inversion between different irreps. It is not known in general whether such band inversion between different irreps in inversion-asymmetric systems
is related to a change in the $Z_2$ topological invariant because it is outside the theory of the symmetry-based indicator and topological quantum chemistry.
In this paper, we study this issue for all the layer groups, and show that the change of the  $Z_2$ topological invariant is directly related to the irreps $(R_{\alpha},R_{\beta})$ exchanged at band inversion at a high-symmetry point (HSP) (see Fig.~\ref{fig:schematic_of_phase_transition}).
We demonstrate its global topology from the effective $\bm{k} \cdot \bm{p}$ theory in the vicinity of one high-symmetry point.

This paper is organized as follows.
Leaving the general theory for later, we first construct a theory of $Z_2$ topological phase transitions under $C_3$ symmetry and time-reversal symmetry in Sec.~\ref{sec:c3}.
We consider an effective theory of four bands of occupied and unoccupied Kramers pairs near the Fermi energy, which follow different irreps of $C_3$ at TRIM.
We show that the $Z_2$ topological invariant always changes when the band gap closes at TRIM and the bands are reversed.
Thus we demonstrate that topological phase transition always occurs at band inversion between different irreps.
In Sec.~\ref{sec:lg}, we apply the method used in Sec.~\ref{sec:c3} to the cases of general layer groups and all the high-symmetry points and list whether topological phase transitions occur.
We find that this problem essentially reduces to a $\bm{k} \cdot \bm{p}$ band theory in five point groups: $C_3, C_4, C_6, S_4,$ and $C_{3h}$, and show that in all the cases, the change of the $Z_2$ topological invariant is entirely determined by the irreps exchanged at the band inversion.
These results are nontrivial because these cases are outside of the theories of symmetry-based indicators and topological quantum chemistry~\cite{Po2017-mk,Kruthoff2017-zh,Bradlyn2017-ot,Bradlyn2018-eu,Khalaf-PRX}.
In Sec.~\ref{sec:discussion}, we conclude the paper.
Throughout this paper, we consider a spinful system with spin--orbit coupling preserving time-reversal symmetry.

\section{\texorpdfstring{$Z_2$}{Z2} Topological phase transitions in a system with \texorpdfstring{$C_3$}{C3} symmetry}\label{sec:c3}
We start with a simple model before constructing our theory on general 2D $Z_2$ topological phase transitions in the next section.
We consider a 2D model with a tunable parameter $m$ with time-reversal symmetry and $C_3$ symmetry, and we assume that the band gap is closed at a TRIM with $C_3$ symmetry at a particular value of the parameter $m$.
Such a gap closing at a TRIM occurs only when the irreps of the valence and the conduction bands at the TRIM are inverted (see Fig.~\ref{fig:schematic_of_phase_transition}).
We study whether the $Z_2$ topological phase transition occurs when the band gap is closed at TRIM with band inversion between different irreps of $C_3$. We take the $\Gamma$ point (${\bf k}=0$) as the TRIM.

\subsection{Setup of the model}\label{sec:setup}
Under the $C_3$ and time-reversal symmetries, there are two possible irreps, $\bar{\Gamma}_4 \bar{\Gamma}_4$ with $C_3 = -1$, and $\bar{\Gamma}_5 \bar{\Gamma}_6$ with $C_3 = e^{\pm \pi i / 3}$, both of which are doubly degenerate due to time-reversal symmetry.
Here, we follow the notation of the irreps in the Bilbao Crystallographic Server~\cite{bilbao,AroyoPerezMato,Aroyo}.
Therefore, we consider a band inversion between the irreps $\bar{\Gamma}_4 \bar{\Gamma}_4$ and $\bar{\Gamma}_5 \bar{\Gamma}_6$, and so we study a four-band model with these irreps at $\Gamma$.
We assume that the system is gapped when $\bm{k} \neq 0$, and the Fermi energy is set between the two Kramers doublets at $\Gamma$.
The effective Hamiltonian of the system in the vicinity of $\bm{k} = 0$ under $C_3$ and time-reversal symmetries can be expressed up to the first order of $\bm{k}$ as
\begin{equation}
    H(\bm{k}, m) =
    \begin{pmatrix}
        m              & \alpha_1 k_-    & \alpha^*_2 k_+ & \alpha^*_3 k_+ \\
        \alpha^*_1 k_+ & m               & \alpha_3 k_-   & -\alpha_2 k_-  \\
        \alpha_2 k_-   & \alpha^*_3 k_+  & -m             & 0              \\
        \alpha_3 k_-   & -\alpha^*_2 k_+ & 0              & -m             \\
    \end{pmatrix},
    \label{eq:hamiltonian_c3}
\end{equation}
where $k_{\pm} \equiv k_x \pm i k_y,\, \alpha_1 = v_1 + i v_2, \,\alpha_2 = v_3 + i v_4,\, \alpha_3 = v_5 + i v_6$ and $m, v_1, v_2, \cdots, v_6$ are real constants.
This Hamiltonian is shown in the basis in the order $(\bar{\Gamma}_5, \bar{\Gamma}_6, \bar{\Gamma}_4, \bar{\Gamma}_4)$ where the $C_3$ rotation is expressed as $C_3 = \mathrm{diag}(e^{\pi i / 3}, e^{-\pi i / 3}, -1, -1)$.
For the purpose of explaining our method simply, we assume $v_2 = v_3 = v_4 = v_6 = 0$ for the moment; we will explain later that the same method is applicable for general cases with nonzero $v_j$.
From this Hamiltonian~\eqref{eq:hamiltonian_c3}, the eigenenergies and eigenstates can be derived as
\begin{align}
    E_{st}(\bm{k}, m)    & = \frac12 \qty(s v_1 k + t \sqrt{(s v_1 k + 2m)^2 + 4 v_5^2 k^2}), \label{eq:E_st_v1_and_v5} \\
    \Phi^{st}(\bm{k}, m) & = \mqty((m + E_{st}(\bm{k}, m))k_{-}                                                         \\
    s (m + E_{st}(\bm{k}, m)) k                                                                                         \\
    s v_5 k k_{+}                                                                                                       \\
    v_5 k_{-}^2),
    \label{eq:eigenstates}
\end{align}
where $s$ and $t$ take the values of $\pm$.
These eigenenergies constantly satisfy $E_{--} \leq E_{+-} \leq E_{-+} \leq E_{++}$ near $\bm{k} = 0$.
For our purpose, we assume that the two lower bands are occupied.

The band gap closes at $\bm{k} = 0$ when $m = 0$ and the $\bar{\Gamma}_4 \bar{\Gamma}_4$ bands and $\bar{\Gamma}_5 \bar{\Gamma}_6$ bands are inverted across $m = 0$.
We consider whether the $Z_2$ topological invariant changes when the parameter $m$ changes from $m = -\delta$ to $m = +\delta$ ($\delta$: positive infinitesimal) as shown in Fig.~\ref{fig:z2-1}(a).

\subsection{Calculation of the $Z_2$ topological invariant}
\begin{figure}[tbp]
    \centering
    \includegraphics[width = 1\columnwidth]{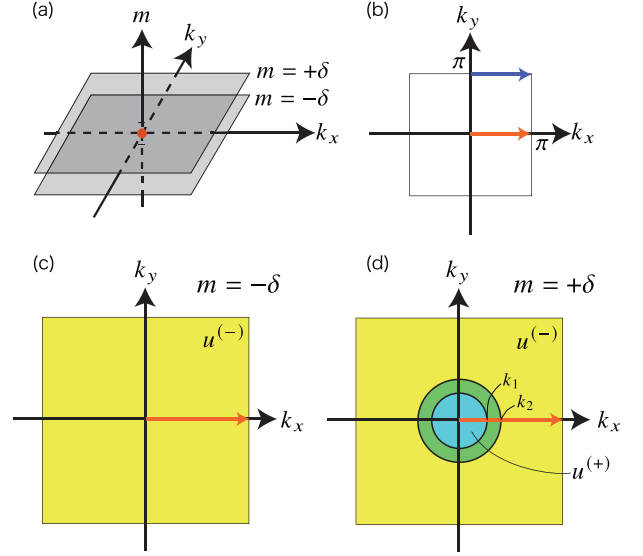}
    \caption{Calculation of the change of the $Z_2$ topological invariant $\nu$ across $m = 0$. (a) $Z_2$ topological invariants for $m = +\delta$ and $m = -\delta$ expressed as integrals on the grey planes in the $kx$--$k_y$--$m$ space. (b) Integral path of the formula of the $Z_2$ topological invariant $\nu$ in Eq.~\eqref{eq:P_theta} with respect to $k_x$ along $k_y = 0$ (red arrow) and $k_y = \pi$ (blue arrow). (c) Eigenstates $v_\alpha (\bm{k}, m)$ for $m < 0$. We take $u^{(-)}$ throughout the Brillouin zone. (d) Eigenstates $v_\alpha (\bm{k}, m)$ for $m > 0$. The eigenstates are defined in each concentric region on the Brillouin zone. In the vicinity of $\bm{k} = 0$ (blue), we take $u^{(+)}$, and in the region where $k$ is large (yellow), we take $u^{(-)}$. Between these regions (green), the states are smoothly connected from $u^{(+)}$ to $u^{(-)}$.}
    \label{fig:z2-1}
\end{figure}

The $Z_2$ topological invariant $\nu$ characterizes topological phases protected by time-reversal symmetry, and $\nu = 1$ and $0$ represent a topological insulator and a normal insulator phase, respectively.
In this study, we use the following formula for the $Z_2$ topological invariant $\nu$ defined in the $k_x$--$k_y$ plane in Ref.~\cite{Fu2006-so}:
\begin{align}
    \nu & = P_\theta^{k_y = \pi} -  P_\theta^{k_y = 0} \quad \mod 2,
    \label{eq:z2_invariant}                                          \\
    \begin{split}
        P_{\theta}^{k_y} & = \frac{1}{2 \pi i} \left[ \int_{0}^{\pi} \dd{k_x} \grad_{k_x} \log \det [w(\bm{k})] \right. \\
            & \hspace{3em} \left. - 2 \log \qty(\frac{\mathrm{Pf}[w(\pi, k_y)]}{\mathrm{Pf}[w(0, k_y)]}) \right] \quad (k_y = 0, \pi).
    \end{split}
    \label{eq:P_theta}
\end{align}
In Eq.~\eqref{eq:P_theta} we perform the integration along $k_y = 0$ or $k_y = \pi$ as shown in Fig.~\ref{fig:z2-1}(b).
Here the $2N \times 2N$ matrix $w$ is expressed using the time-reversal operator $\Theta$ and the Bloch wavefunction $u_{\bm{k}, n}$ of the $n$th band of the system as
\begin{equation}
    w_{mn}(\bm{k}) = \mel{u_{-\bm{k}, m}}{\Theta}{u_{\bm{k},n}},
    \label{eq:w_matrix}
\end{equation}
where $2N$ is the number of occupied bands and $m$ and $n$ run over the band indices of the $2 N$ occupied bands.
Here, $P_\theta^{k_y = 0}$ and $P_\theta^{k_y = \pi}$ are integers defined in terms of modulo 2.
In particular, when the $2N$ wavefunctions are classified into two groups,  $u_\alpha^{\mathrm{I}} (\bm{k})$ and $u_\alpha^{\mathrm{I\hspace{-.01em}I}}(\bm{k})$ $(\alpha = 1, 2, \cdots, N)$, which satisfy
\begin{equation}
    \begin{split}
        \ket{u^{\mathrm{I}}_{\alpha}(-\bm{k})}  & = \Theta \ket{u^{\mathrm{I\hspace{-.01em}I}}_{\alpha}(\bm{k})},  \\
        \ket{u^{\mathrm{I\hspace{-.01em}I}}_{\alpha}(-\bm{k})} & = -\Theta \ket{u^{\mathrm{I}}_{\alpha}(\bm{k})},
    \end{split}
    \label{eq:gauge}
\end{equation}
the matrix $w$ becomes $i \sigma_y \otimes I_N$ ($\sigma_y$: Pauli matrix, $I_N: N \times N$ identity matrix), and we get a simple result $\mathrm{Pf} [w(\bm{k})] = 1$ and $\det [w(\bm{k})] = 1$.
Thus, when we can take the gauge of the wavefunctions to satisfy Eq.~\eqref{eq:gauge} over the entire Brillouin zone, $P_\theta^{k_y = 0} = 0$ and $P_\theta^{k_y = \pi} = 0$ holds, and the $Z_2$ topological invariant $\nu$ is trivial, as shown in Ref.~\cite{Fu2006-so}.
Namely, when $\nu$ is nontrivial $(\nu = 1)$, one cannot choose a gauge that satisfies Eq.~\eqref{eq:gauge} in the whole Brillouin zone.

Our purpose is to calculate the jump of the topological invariant $\nu$ across $m=0$, but now
we encounter a problem: the eigenstates~\eqref{eq:eigenstates} are zero at $\bm{k} = 0$, so they cannot be normalized. To overcome this problem,
we need to construct linear combinations of the occupied eigenstates~\eqref{eq:eigenstates}, which have finite nonzero limits at $m\rightarrow 0$.
We note that near ${\bf k}=0$, this procedure should be done in a different way between the two limits $m\rightarrow +0$ and $m\rightarrow -0$, and it involves some technical and lengthy
calculations, as we show in the following. The readers can skip the calculation details and proceed to the next subsection for the results of the calculation.

We note that since the effective Hamiltonian~\eqref{eq:hamiltonian_c3} gives $H(\bm{k} = 0, m) = \mathrm{diag}(m, m, -m, -m)$, the occupied states at $\bm{k} = 0$ are
\begin{equation}
    \begin{split}
        \mqty(0 \\ 0\\ 1\\ 0), \mqty(0\\ 0\\ 0\\ 1) \qquad (m > 0), \\
        \mqty(1 \\ 0\\ 0\\ 0), \mqty(0\\ 1\\ 0\\ 0) \qquad (m < 0),
    \end{split}
    \label{eq:occ_states_0}
\end{equation}
with a discontinuity at $m = 0$.
Thus we need to construct eigenstates which reduce to Eq.~\eqref{eq:occ_states_0} for $\bm{k} \to 0$.
Let $u_{1}^{(+)}(\bm{k}, m), u_{2}^{(+)}(\bm{k}, m)$ be the orthonormal occupied states in the region $m > 0$, and $u_{1}^{(-)}(\bm{k}, m), u_{2}^{(-)}(\bm{k}, m)$ in the region $m < 0$ as well.
We impose a condition that their behavior in $\bm{k} \to 0$ are
\begin{equation}
    \begin{split}
        u_{1}^{(+)}(\bm{k}, m) \to \mqty(0 \\ 0\\ 1\\ 0), \, u_{2}^{(+)}(\bm{k}, m) \to \mqty(0 \\ 0\\ 0\\ 1) \qquad (m = + \delta), \\
        u_{1}^{(-)}(\bm{k}, m) \to \mqty(1 \\ 0\\ 0\\ 0), \, u_{2}^{(-)}(\bm{k}, m) \to \mqty(0 \\ 1\\ 0\\ 0) \qquad (m = - \delta).
    \end{split}
    \label{eq:occ_states_near_k0}
\end{equation}
It is noted that $u_{\alpha}^{(\pm)}(\bm{k}, m) \quad (\alpha = 1, 2)$ are not necessarily eigenvectors of the Hamiltonian but are linear combinations of eigenvectors for occupied states in general~\cite{Fu2006-so}.
As we discuss later, we can construct $u_\alpha^{(\pm)}(\bm{k}, m)$ from Eq.~\eqref{eq:hamiltonian_c3}.

\begin{figure}[tbp]
    \centering
    \includegraphics[width = 1\columnwidth]{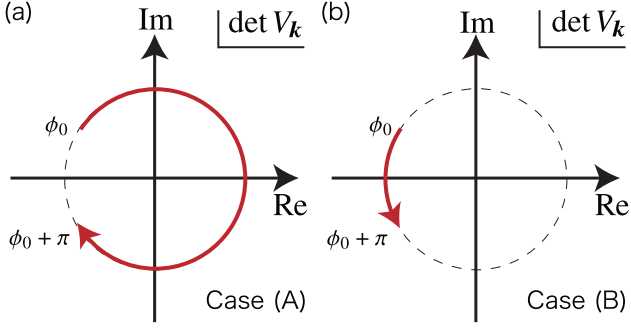}
    \caption{Two cases for the change of the $Z_2$ topological invariant $\nu$ across $m = 0$. Whether $Z_2$ topological phase transition occurs depends on whether $\det V_{\bm{k}}$ passes through the positive side of the real axis (a) or the negative side (b) at odd times when the azimuth angle of $\bm{k}$ is changed from $\phi_0$ to $\phi_0 + \pi$. In Case (B), a $Z_2$ topological phase transition occurs, while not in Case (A).}
    \label{fig:z2-2}
\end{figure}

Now we calculate the change of the $Z_2$ topological invariant when $m$ changes from $- \delta$ to $\delta$.
We have already defined the wavefunctions $u_\alpha^{(\pm)}(\bm{k}, m)$ for $m > 0$ and $m < 0$, but they are still not convenient for our purpose.
Therefore, starting from $u_\alpha^{(\pm)}(\bm{k}, m)$, we construct another form of wavefunctions of the occupied states $v_\alpha (\bm{k}, m)$ near $m = 0$, which are convenient for a calculation of the change of the $Z_2$ topological invariant.
The key for this calculation is that the eigenvectors for a large $|\bm{k}|$ are unaffected by this infinitesimal change of $m$.
Based on this observation, we define $v_\alpha(\bm{k}, m)$ as follows.
When $m = -\delta$, the eigenstates $v_\alpha (\bm{k}, m)$ are equal to $u_\alpha^{(-)} (\bm{k}, m)$ over the entire region, as shown in Fig.~\ref{fig:z2-1}(c).
When $m = +\delta$, the wavefunctions for a large $|\bm{k}|$ are the same as $m = -\delta$ and we take $v_\alpha (\bm{k}, m) = u_\alpha^{(-)} (\bm{k}, m)$ (by smoothly extending the definition of $u_\alpha^{(-)} (\bm{k}, m)$ to $m = +\delta$, which is possible for a large $|\bm{k}|$).
Meanwhile, for those for a small $|\bm{k}|$, we take $v_\alpha (\bm{k}, m) = u_\alpha^{(+)} (\bm{k}, m)$ to account for the jump of eigenstates at $\bm{k} = 0$ and $m = 0$.
Here we impose a condition that $v_\alpha (\bm{k}, m)$ should be continuous functions of $\bm{k}$ to calculate the $Z_2$ topological invariant, and thus we set
\begin{equation}
    v_\alpha (\bm{k}, +\delta) = \left\{ \,
    \begin{aligned}
        u_\alpha^{(+)} (\bm{k}, +\delta) & \quad |\bm{k}| \leq k_1 \\
        u_\alpha^{(-)} (\bm{k}, +\delta) & \quad |\bm{k}| \geq k_2
    \end{aligned}
    \right. ,
    \label{eq:u_alpha_plus_delta}
\end{equation}
where $k_1$ and $k_2$ $(> k_1)$ are positive constants.
Within the region $k_1 \leq |\bm{k}| \leq k_2$, we will define the wavefunctions $v_\alpha (\bm{k}, +\delta)$ later so that $v_\alpha (\bm{k}, +\delta)$ are continuous functions of $\bm{k}$, as shown in Fig.~\ref{fig:z2-1}(d).

Using this wavefunction $v_\alpha(\bm{k}, m)$, we can calculate the change of the $Z_2$ topological invariant $\nu$ via Eqs.~\eqref{eq:z2_invariant} and \eqref{eq:P_theta}.
For this purpose, we choose the gauge of $u^{(\pm)} (\bm{k}, m)$ to satisfy Eq.~\eqref{eq:gauge}.
By this gauge choice, the calculation of the $Z_2$ topological invariant $\nu$ becomes simpler; for the wavefunction in Eq.~\eqref{eq:u_alpha_plus_delta}, only the wavefunctions in the region $k_1 < |\bm{k}| < k_2$ contribute to $\nu$.
Detailed calculations of $v_\alpha (\bm{k}, m)$ and of the change of the $Z_2$ topological invariant are described in Appendix~\ref{sec:detailed}.
As a result of the lengthy calculation, we conclude that the cases are classified into two cases.
Let us define a unitary matrix $V_{\bm{k}}$ by
\begin{equation}
    (u_1^{(+)}(\bm{k}, +\delta), u_2^{(+)}(\bm{k}, +\delta)) = (u^{(-)}_1(\bm{k}, +\delta), u^{(-)}_2(\bm{k}, +\delta)) V_{\bm{k}},
    \label{eq:def_V_k}
\end{equation}
and we consider the change of $\det V_{\bm{k}}$ when $\bm{k}$ moves along the circle $\bm{k} = (k_1 \cos \phi, k_1 \sin \phi)$.
Then as we change $\phi$ from $\phi_0$ to $\phi_0 + \pi$, where $\phi_0$ is a constant, $\det V_{\bm{k}}$ changes from $\det V_{\phi_0}$ to its complex conjugate, $\det V_{\phi_0 + \pi} \equiv (\det V_{\phi_0})^*$ along the unit circle, which follows from time-reversal symmetry.
Then, its trajectory crosses either the positive or negative part of the real axis an odd number of times (see Fig.~\ref{fig:z2-2}), which we classify into cases (A) and (B), respectively.
Then, through the calculation in Appendix~\ref{sec:detailed}, we get
\begin{equation}
    \nu(m = +\delta) - \nu(m = -\delta) =
    \left\{
    \begin{aligned}
         & 0 \qq{: case (A)}  \\
         & 1 \qq{: case (B)}.
    \end{aligned}
    \right.
    \label{eq:nu-nu_case}
\end{equation}

Thus, we have established how to calculate the change of $\nu$ across the gap-closing point $m = 0$.
Now we calculate the $Z_2$ topological invariant for the present case of $C_3$ symmetry.
The eigenstates for the occupied states, $\Phi^{+-}(\bm{k}, m)$ and $\Phi^{--}(\bm{k}, m)$, are given by Eq.~\eqref{eq:eigenstates}.
Nevertheless, they are not normalizable at $\bm{k} = 0$, and thus they do not satisfy Eq.~\eqref{eq:occ_states_near_k0}.
To construct the occupied states $u_\alpha^{(\pm)}(\bm{k})$ that satisfy Eq.~\eqref{eq:occ_states_near_k0}, we need to make linear combinations.
To make linear combinations, we define $2 \times 2$ matrices $A$ and $B$ as
\begin{equation}
    M = (\Phi_{\bm{k}}^{+-}, \Phi_{\bm{k}}^{--}) = \mqty(A \\ B),
    \label{eq:def_M}
\end{equation}
and in the particular case of $v_2 = v_3 = v_4 = v_6 = 0$ it is given by
\begin{equation}
    M = \begin{pmatrix}(m + E_{+-})k_{-} & (m + E_{--})k_{-}\\ k (m + E_{+-}) & -k (m + E_{--})\\ v_5 k k_{+} & -v_5 k k_{+}\\ v_5 k_{-}^2 & v_5 k_{-}^2\end{pmatrix} = \mqty(A\\ B)
\end{equation}
By definition (Eq.~\eqref{eq:occ_states_near_k0}), it follows that $(u_1^{(+)}, u_2^{(+)}) = M B^{-1}$ and $(u_1^{(-)}, u_2^{(-)}) = M A^{-1}$ hold asymptotically when $\bm{k}\to 0$, and we obtain $V_{\bm{k}} = A B^{-1}$ for $\bm{k} \sim 0$.
Through a direct calculation, we obtain
\begin{equation}
    \begin{split}
        \det V_{\bm{k}} = - \frac{(m + E_{+-})(m + E_{--})}{k^2 \sum_{i = 1}^6 v_i^2} \to -1 \\
        (\bm{k} \to 0, m \to 0)
        \label{eq:calc_det_V_k}\\
    \end{split},
\end{equation}
where $E_{st}$ is the eigenvalue~\eqref{eq:E_st_v1_and_v5} for general values of $v_1, v_2, \cdots, v_6$, and is expressed by
\begin{equation}
    \begin{split}
        E_{st} = \frac{1}{2} & \left[s \sqrt{v_1^2 + v_2^2} k \right.
            \\& \left. + t \sqrt{(s \sqrt{v_1^2 + v_2^2} k + 2m)^2 + 4 k^2 \sum_{i = 3}^6 v_i^2} \right].
    \end{split}
\end{equation}
From Eq.~\eqref{eq:calc_det_V_k}, it is the case (B), and the change in the $Z_2$ topological invariant across $m = 0$ is 1.
In other words, if the system has $C_3$ symmetry, the $Z_2$ topological phase transition always occurs when the band gap is closed at TRIM, and the $\bar{\Gamma}_4 \bar{\Gamma}_4$ bands and the $\bar{\Gamma}_5 \bar{\Gamma}_6$ bands are inverted.

To confirm this result, in Appendix~\ref{sec:TBmodel}, we present a calculation on a tight-binding model with $C_3$- and time-reversal symmetries, with band inversion at the $\Gamma$ point by changing a parameter.
We show that the $Z_2$ topological invariant changes at the band inversion, which confirms our calculation.

\subsection{Summary of Sec.~\ref{sec:c3}}
Thus we have shown that in systems with $C_3$ and time-reversal symmetries, the band inversion between
the different $C_3$ doublets, \textit{i}.\textit{e}.,
$\bar{\Gamma}_4 \bar{\Gamma}_4$ ($C_3=-1$) and  $\bar{\Gamma}_5 \bar{\Gamma}_6$ ($C_3=e^{\pm \pi i/3}$)
is always accompanied by the change of the $Z_2$ topological invariant. The same conclusion is obtained in Ref.~\cite{Yu2020-ik} by continuous deformation of the system to an inversion-symmetric system. We discuss the relationship between our result and the result in Ref.~\cite{Yu2020-ik} in
Sec.~\ref{sec:discussion}.

\section{\texorpdfstring{$Z_2$}{Z2} Topological phase transitions with layer-group symmetries}\label{sec:lg}
In this section, we construct a theory of the 2D $Z_2$ topological phase transition under general layer group symmetry other than the $C_3$ symmetry studied in the previous section.
In the following, we consider possible phase transitions at all the high symmetry points in all the 80 layer groups.
Among the 80 layer groups, the 31 layer groups contain inversion symmetry, and the $Z_2$ topological invariant $\nu$ is expressed in terms of parity eigenvalues at TRIM.
Thus the $Z_2$ topological phase transition is always accompanied by a band inversion at a TRIM, where the parity eigenvalues are switched.

Thus we focus on the remaining 49 layer groups, not containing inversion symmetry.
By our detailed analysis in Sec.~\ref{sec:topological_phase_transitions_in_80_layer_group_symmetries}, we find that the local point-group symmetries at the HSPs to be considered are limited to five cases: rotation symmetries $C_3, C_4, C_6,$ and rotoinversion symmetries $S_4, C_{3h}$.
Therefore, in the following, we consider a model of a 2D topological insulator with time-reversal symmetry and each of the point group symmetries $C_3, C_4, C_6, S_4,$ and $C_{3h}$, and discuss whether the $Z_2$ topological phase transition occurs when the band gap is closed at an HSP.
The case with $C_3$ is already discussed in Sec.~\ref{sec:c3}, so we study the other cases in Sec.~\ref{sec:4}--\ref{sec:c3h}.

\subsection{Topological phase transitions in 80 layer-group symmetries}\label{sec:topological_phase_transitions_in_80_layer_group_symmetries}
In this subsection, we consider whether the band-gap closing at HSPs is accompanied by a $Z_2$ topological phase transition in each of the 80 layer groups listed in Table \ref{tb:80LGs}.
Among the HSPs, some are non-TRIM.
Such non-TRIM points are limited to $K$ and $K'$ points in trigonal and hexagonal crystal systems, and we can show that band inversions at $K$ and $K'$ points are always accompanied by a change of the $Z_2$ topological invariant by following Ref.~\cite{Murakami2007-lt}.
In Sec.~\ref{sec:setup} of Ref.~\cite{Murakami2007-lt}, the gap closings at the non-TRIM
points in inversion asymmetric systems are studied, and it is concluded that such gap closings are always accompanied by
$Z_2$ topological phase transition, where only the time-reversal symmetry is assumed.
Meanwhile, in the present case, additional crystallographic symmetries such as threefold rotational symmetry are assumed, but
the Hamiltonian around the gap-closing points (at $K$ and $K'$) is still a 2D massive Dirac Hamiltonian as described in Ref.~\cite{Murakami2007-lt},
and so the conclusion remains the same.
Thus in the following, we restrict ourselves to band inversions at TRIM.

First, among the 80 layer groups, the 31 layer groups (marked ``inv'' in Table \ref{tb:80LGs}) have inversion symmetries. In these 31 layer groups, the
$Z_2$ topological invariant is written in terms of parity eigenvalues at TRIM \cite{FuKane-inversion}, which means that their $Z_2$ topological phase transitions
are well understood. Namely, when the band gap closes
at TRIM, and thereby the parity eigenvalues are switched between the occupied and unoccupied bands, it is always accompanied by the $Z_2$ topological
phase transition. Thus we need to study the remaining 49 layer groups and examine whether an insulator-to-insulator transition
accompanies a $Z_2$ topological phase transition.

\begin{figure}[tbp]
    \centering
    \includegraphics[width = 1\columnwidth]{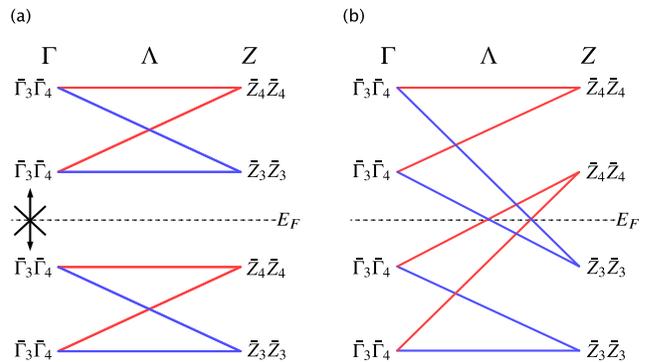}
    \caption{Hourglass band structure in the layer group No. 9 ($p2_1 11$) along the $\Gamma$--$Z$ line, where the values of $\bm{k}$ are $\Gamma(000)$, $Z(0\pi 0)$, and $\Lambda(0 k_y 0)$. The screw eigenvalues are $-i$ in $\bar{\Gamma}_3$, $i$ in $\bar{\Gamma}_4$, $1$ in $\bar{Z}_3$, $-1$ in $\bar{Z}_4$, $i e^{-i k_y / 2}$ in $\bar{\Lambda}_3$ (blue) and $i e^{i k_y / 2}$ in $\bar{\Lambda}_4$ (red). $(\bar{\Gamma}_3, \bar{\Gamma}_4)$, $(\bar{Z}_3, \bar{Z}_3)$, $(\bar{Z}_4, \bar{Z}_4)$ are Kramers pairs. (a) Gapped band structure. The band gap cannot be inverted at $\Gamma$ because there is only one possible irrep, $\bar{\Gamma}_3 \bar{\Gamma}_4$, and the band gap cannot close. Meanwhile, the gap can be switched at $Z$, leading to the band structure in (b), where the gap does not open because different irreps $\bar{\Lambda}_3$ and $\bar{\Lambda}_4$ must cross.}
    \label{fig:hourglass_GM-Z}
\end{figure}

For a further study on $Z_2$ topological phase transitions, we find that among the 49 layer groups, we can exclude 28 layer groups for either of the following two reasons:
\renewcommand{\theenumi}{(\roman{enumi})}
\begin{enumerate}
    \item The TRIM considered has only one irreducible representation. Therefore, the band gap cannot close at that point by tuning a single parameter due to level repulsion.
    \item The bands along a high-symmetry line between two TRIM always form an hourglass band structure \cite{Young2015-cv,Wieder2016-tf,Bzdusek2016-ud}. Therefore, even when the band gap closes between states with different irreps, a gap does not open again. An example is shown in Fig.~\ref{fig:hourglass_GM-Z} for layer group No. 9 ($p2_1 11$), where we show that the band inversion at the $Z$ point does not open a gap.
\end{enumerate}
In the case (i), if we construct an effective model around the TRIM, there will be various off-block-diagonal elements that
mix the valence and the conduction bands, as discussed in Ref.~\cite{Murakami2007-lt} for the simplest case. These off-block-diagonal elements give rise to level repulsion at the HSP, and it is impossible to make all these elements zero by tuning only one parameter. Thus the band gap never closes. On the other hand, the case (ii) occurs
only in some nonsymmorphic layer groups. At every TRIM, all the states are doubly degenerate due to Kramers theorem, and in some nonsymmorphic layer groups,
the Kramers doublets should always switch partners between two TRIM due to the $\bm{k}$-dependence of the eigenvalues for nonsymmorphic symmetry operations~\cite{hourglass}.
This prohibits an insulator-to-insulator transition by changing one parameter, as exemplified in Fig.~\ref{fig:hourglass_GM-Z}.
In Table \ref{tb:80LGs}, we show how these cases (i) and (ii) appear in the 80 layer groups.
In the table, the layer groups with ``(i)" means that (i) applies to all the TRIM in that layer group. Similarly, ``(ii)" means that (ii) applies to all the TRIM.
Moreover, ``(i)(ii)" means that at all the TRIM, either (i) or (ii) applies. The blank entities in the table mean that at some of the TRIM, neither (i) nor (ii) applies, and these cases are the targets of our theory.

\begin{table*}[htbp]
    \centering
    \begin{tabular}{lllllllllllllll}
        No.   & Symbol          & Type \ \ \    & No.    & Symbol            & Type   \ \ \  & No.    & Symbol              & Type  \ \ \  & No.    & Symbol            & Type  \ \ \  &
        No.   & Symbol          & Type   \ \ \                                                                                                                                                                              \\
        \hline
        $1$   & $ p 1 $         & (i)           & $ 2 $  & $ p \bar{1} $     & inv           & $ 3 $  & $ p 1 1 2 $         & (i)          & $ 4 $  & $ p 1 1 m $       & (i)          & $ 5 $  & $ p 1 1 a  $       & (ii) \\
        $6$   & $ p 1 1 2/m $   & inv           & $ 7 $  & $ p 1 1 2/a $     & inv           & $ 8 $  & $ p 2 1 1 $         & (i)          & $ 9 $  & $ p 2_1 1 1 $     & (ii)         & $ 10 $ & $ c 2 1 1  $       & (i)  \\
        $11$  & $ p m 1 1 $     & (i)           & $ 12 $ & $ p b 1 1 $       & (ii)          & $ 13 $ & $ c m 1 1 $         & (i)          & $ 14 $ & $ p 2/m 1 1 $     & inv          & $ 15 $ & $ p 2_1/m 1 1  $   & inv  \\
        $16$  & $ p 2/b11 $     & inv           & $ 17 $ & $ p 2_1/b 1 1 $   & inv           & $ 18 $ & $ c 2/m 1 1 $       & inv          & $ 19 $ & $ p 2 2 2 $       & (i)          & $ 20 $ & $ p 2_1 2 2  $     & (ii) \\
        $21$  & $ p 2_1 2_1 2 $ & (ii)          & $ 22 $ & $ c 2 2 2 $       & (i)           & $ 23 $ & $ p m m 2 $         & (i)          & $ 24 $ & $ p m a 2 $       & (ii)         & $ 25 $ & $ p b a 2  $       & (ii) \\
        $26$  & $ c m m 2 $     & (i)           & $ 27 $ & $ p m 2 m $       & (i)           & $ 28 $ & $ p m 2_1b $        & (i)          & $ 29 $ & $ p b 2_1m $      & (i)          & $ 30 $ & $ p b 2 b  $       & (i)  \\
        $31$  & $ p m 2 a $     & (ii)          & $ 32 $ & $ p m 2_1n $      & (i)(ii)       & $ 33 $ & $ p b 2_1a $        & (i)(ii)      & $ 34 $ & $ p b 2 n $       & (i)(ii)      & $ 35 $ & $ c m 2 m  $       & (i)  \\
        $36$  & $ c m 2 e $     & (ii)          & $ 37 $ & $ p m m m $       & inv           & $ 38 $ & $ p m a a $         & inv          & $ 39 $ & $ p b a n $       & inv          & $ 40 $ & $ p m a m  $       & inv  \\
        $41$  & $ p m m a $     & inv           & $ 42 $ & $ p m a n $       & inv           & $ 43 $ & $ p b a a $         & inv          & $ 44 $ & $ p b a m $       & inv          & $ 45 $ & $ p b m a  $       & inv  \\
        $46$  & $ p m m n $     & inv           & $ 47 $ & $ c m m m $       & inv           & $ 48 $ & $ c m m e $         & inv          & $ 49 $ & $ p 4 $           &              & $ 50 $ & $ p \bar{4}  $     &      \\
        $51 $ & $ p 4/m $       & inv           & $ 52 $ & $ p 4/n $         & inv           & $ 53 $ & $ p 4 2 2 $         &              & $ 54 $ & $ p 4 2_1 2 $     &              & $ 55 $ & $ p 4 m m  $       &      \\
        $56$  & $ p 4 b m $     &               & $ 57 $ & $ p \bar{4} 2 m $ &               & $ 58 $ & $ p \bar{4} 2_1 m $ &              & $ 59 $ & $ p \bar{4} m 2 $ &              & $ 60 $ & $ p \bar{4} b 2  $ &      \\
        $61$  & $ p 4/m m m $   & inv           & $ 62 $ & $ p 4/n b m $     & inv           & $ 63 $ & $ p 4/m b m $       & inv          & $ 64 $ & $ p 4/n m m $     & inv          & $ 65 $ & $ p 3  $           &      \\
        $66 $ & $ p \bar{3} $   & inv           & $ 67 $ & $ p 3 1 2 $       &               & $ 68 $ & $ p 3 2 1 $         &              & $ 69 $ & $ p 3 m 1 $       &              & $ 70 $ & $ p 3 1 m  $       &      \\   $71 $ & $ p \bar{3} 1 m $  & inv& $ 72 $ & $ p \bar{3} m 1 $  & inv& $ 73 $ & $ p 6 $ && $ 74 $ & $ p \bar{6} $ && $ 75 $ & $ p 6/m  $  & inv\\
        $76 $ & $ p 6 2 2 $     &               & $ 77 $ & $ p 6 m m $       &               & $ 78 $ & $ p \bar{6} m 2 $   &              & $ 79 $ & $ p \bar{6} 2 m $ &              & $ 80 $ & $ p 6/m m m  $     & inv  \\ \hline
    \end{tabular}
    \caption{80 layer groups and their classifications in terms of the band-gap closing. Here the layer group numbers and symbols are shown, together with its types discussed in the text. ``inv'' means that the layer group contains the inversion symmetry, and therefore the $Z_2$ topological phase transition is always accompanied by a band inversion involving the exchange of parities at TRIM. ``(i)'' means that
        all the TRIM have only one irreducible representation; therefore, the band gap cannot close at any of the high-symmetry points. ``(ii)'' means that all the TRIM mutually form
        an hourglass band structure. ``(i)(ii)'' means that (i) applies to some of the TRIM and (ii) applies to the rest of the TRIM. As a result, in the layer groups with (i), (ii), an insulator-to-insulator transition never occurs.
        The cases with a blank in the ``Type'' section are the ones we study here, where an insulator-to-insulator transition occurs at TRIM.}
    \label{tb:80LGs}\end{table*}

Therefore, we are left with 21 layer groups listed in Table \ref{tb:17LGs} with a blank in the ``Type" column in Table~\ref{tb:80LGs}. Some of the TRIM allow band inversion between different irreps at an insulator-to-insulator transition in these layer groups.
Such TRIM and the point-group symmetries at these points are
also listed in Table \ref{tb:17LGs}. We see varieties of point group symmetries, but the irreps in these cases are essentially the same as those
in one of the subgroups of the point group (as shown in the column ``Subgroup'' in Table \ref{tb:17LGs}). Thus, we need to consider only the Hamiltonians around the HSPs,
having the point-group symmetry $C_3,C_4,C_6,S_4,$ or $C_{3h}$.
From Table~\ref{tb:17LGs}, we see that the TRIM to be studied is $\Gamma$ and $M$.
We have looked at the case with threefold rotational symmetry $C_3$ in the previous section, and we study the remaining cases in the following subsections (as summarized in Table~\ref{tb:PGs}).

\begin{table}[htbp]
    \centering
    \begin{tabular}{lllll}
        No. & Symbol             & HSPs       & Point group & Subgroup \\ \hline
        49  & $ p 4 $            & $\Gamma M$ & $C_4$       & $C_4$    \\
        50  & $ p \bar{4}  $     & $\Gamma M$ & $S_4$       & $S_4$    \\
        53  & $ p 4 2 2 $        & $\Gamma M$ & $D_4$       & $C_4$    \\
        54  & $ p 4 2_1 2$       & $\Gamma $  & $D_4$       & $C_4$    \\
        55  & $ p 4 m m  $       & $\Gamma M$ & $C_{4v}$    & $C_4$    \\
        56  & $ p 4 b m$         & $\Gamma$   & $C_{4v}$    & $C_4$    \\
        57  & $ p \bar{4} 2 m $  & $\Gamma M$ & $D_{2d}$    & $S_4$    \\
        58  & $ p \bar{4} 2_1 m$ & $\Gamma$   & $D_{2d}$    & $S_4$    \\
        59  & $ p \bar{4} m 2 $  & $\Gamma M$ & $D_{2d}$    & $S_4$    \\
        60  & $ p \bar{4} b 2$   & $\Gamma$   & $D_{2d}$    & $S_4$    \\
        65  & $ p 3  $           & $\Gamma $  & $C_3$       & $C_3$    \\
        67  & $ p 3 1 2 $        & $\Gamma $  & $D_3$       & $C_3$    \\
        68  & $ p 3 2 1 $        & $\Gamma $  & $D_3$       & $C_3$    \\
        69  & $ p 3 m 1 $        & $\Gamma $  & $C_{3v}$    & $C_3$    \\
        70  & $ p 3 1 m  $       & $\Gamma $  & $C_{3v}$    & $C_3$    \\
        73  & $ p 6 $            & $\Gamma $  & $C_6$       & $C_6$    \\
        74  & $ p \bar{6} $      & $\Gamma $  & $C_{3h}$    & $C_{3h}$ \\
        76  & $ p 6 2 2 $        & $\Gamma $  & $D_6$       & $C_6$    \\
        77  & $ p 6 m m $        & $\Gamma $  & $C_{6v}$    & $C_6$    \\
        78  & $ p \bar{6} m 2 $  & $\Gamma $  & $D_{3h}$    & $C_{3h}$ \\
        79  & $ p \bar{6} 2 m $  & $\Gamma $  & $D_{3h}$    & $C_{3h}$ \\ \hline
    \end{tabular}
    \caption{21 Layer groups where a band inversion between different irreps at high-symmetry points is
        possible. We only show the layer groups without inversion symmetry. The 31 layer groups with inversion symmetry are excluded since its topological phase transition is already well understood. }  \label{tb:17LGs}\end{table}

\subsection{\texorpdfstring{$C_4$}{C4} symmetry}\label{sec:4}
We study a model with point group $C_4$ as a local symmetry at a TRIM and discuss whether the $Z_2$ topological phase transition occurs when the band gap is closed at the TRIM.
Under the $C_4$ and time-reversal symmetries, there are two irreps, $\bar{\Gamma}_6 \bar{\Gamma}_8$ with $C_4 = e^{\pm \pi i / 4}$ and $\bar{\Gamma}_5 \bar{\Gamma}_7$ with $C_4 = e^{\pm 3 \pi i / 4}$, both of which are Kramers doublets.
Thus, we need to consider a band inversion between the irreps $\bar{\Gamma}_6 \bar{\Gamma}_8$ and $\bar{\Gamma}_5 \bar{\Gamma}_7$.
The $4 \times 4$ effective Hamiltonian in the vicinity of TRIM for this system can be derived using an approximation up to the first order in $\bm{k}$ as follows:
\begin{equation}
    H(\bm{k}) =
    \begin{pmatrix}
        m              & \alpha_1 k_-    & \alpha_2^* k_+ & 0              \\
        \alpha_1^* k_+ & m               & 0              & -\alpha_2 k_-  \\
        \alpha_2 k_-   & 0               & -m             & \alpha_3^* k_+ \\
        0              & -\alpha_2^* k_+ & \alpha_3 k_-   & -m             \\
    \end{pmatrix},
    \label{eq:hamiltonian_c4}
\end{equation}
where $\bm{k}$ is measured from the TRIM with $C_4$ symmetry, $\alpha_1 = v_1 + i v_2, \alpha_2 = v_3 + i v_4, \alpha_3 = v_5 + i v_6$ are complex and $m, v_1, \cdots, v_6$ are real parameters.
Here, $m$ is considered a parameter that drives the closing of the gap.

Now we compute the eigenvectors from the effective Hamiltonian~\eqref{eq:hamiltonian_c4} using the same procedure as in Sec.~\ref{sec:c3}, and then compute Eq.~\eqref{eq:calc_det_V_k} from Eq.~\eqref{eq:def_M}.
By taking $m \to 0$, one can calculate the eigenstates and eigenvalues analytically.
We obtain in the limit of $|\bm{k}| \to +0$
\begin{equation}
    \det V_{\bm{k}} = -1,
    \label{eq:det_V_k_c4}
\end{equation}
which is independent of $\phi (=\arg \bm{k})$.
Therefore, it is in the case (B), and if the system has $C_4$ symmetry, the $Z_2$ topological phase transition always occurs when the band gap closes at the $C_4$-symmetric TRIM.

\subsection{\texorpdfstring{$C_6$}{C6} symmetry}\label{sec:6}
At the TRIM under the $C_6$ symmetry and time-reversal symmetry, there are three irreps, $\bar{\Gamma}_{10} \bar{\Gamma}_{11}$ with $C_6 = e^{\pm \pi i / 6}$, $\bar{\Gamma}_7 \bar{\Gamma}_8$ with $C_6 = e^{\pm  \pi i / 2}$ and $\bar{\Gamma}_9 \bar{\Gamma}_{12}$ with $C_6 = e^{\pm 5 \pi i / 6}$, all of which are doubly degenerate due to time-reversal symmetry.
Thus, we need to consider three cases of band inversions between the pairs of irreps (a)$(\bar{\Gamma}_{10} \bar{\Gamma}_{11}, \bar{\Gamma}_7 \bar{\Gamma}_8)$, (b)$(\bar{\Gamma}_7 \bar{\Gamma}_8, \bar{\Gamma}_9 \bar{\Gamma}_{12})$, and (c)$(\bar{\Gamma}_9 \bar{\Gamma}_{12}, \bar{\Gamma}_{10} \bar{\Gamma}_{11})$.
We then find that in the two cases (a) and (b), the effective Hamiltonian in the vicinity of TRIM using an approximation up to the first order of $\bm{k}$ is the same as the effective Hamiltonian in the case of $C_3$ symmetry ~\eqref{eq:hamiltonian_c3} with $v_5 = v_6 = 0$; it is natural because these two cases (a) and (b) reduce to the one considered in Sec.~\ref{sec:c3} thanks to $(C_6)^2 = C_3$.
Hence, the $Z_2$ phase transition always accompanies band inversions in (a) and (b).

In the case (c), the $4 \times 4$ effective Hamiltonian in the vicinity of TRIM up to the first order of $\bm{k}$ is
\begin{equation}
    H(\bm{k}) =
    \begin{pmatrix}
        m              & \alpha_1 k_- & 0            & 0              \\
        \alpha_1^* k_+ & m            & 0            & 0              \\
        0              & 0            & -m           & \alpha_2^* k_+ \\
        0              & 0            & \alpha_2 k_- & -m             \\
    \end{pmatrix},
    \label{eq:hamiltonian_c6_1}
\end{equation}
where $\alpha_1$ and $\alpha_2$ are complex constants and $\bm{k}$ is measured from the TRIM with $C_6$ symmetry.
Here, $m$ is considered a parameter that drives the closing of the gap.
However, since the band gap of the Hamiltonian~\eqref{eq:hamiltonian_c6_1} permanently closes along the circle with $|\bm{k}| = 2m / (|\alpha_1| + |\alpha_2|)$ other than the TRIM when $m \neq 0$, this Hamiltonian is not appropriate for considering the gap-closing at TRIM.
This is an artifact of truncating up to the linear order in $\bm{k}$.
Therefore, we derive the effective Hamiltonian up to the second order of $\bm{k}$ as follows:
\begin{equation}
    H(\bm{k}) =
    \begin{pmatrix}
        m + u_1 k^2     & \alpha_1 k_-  & \beta_1 k_+^2 & 0               \\
        \alpha_1^* k_+  & m + u_1 k^2   & 0             & \beta_1^* k_-^2 \\
        \beta_1^* k_-^2 & 0             & -m + u_2 k^2  & \alpha_2^* k_+  \\
        0               & \beta_1 k_+^2 & \alpha_2 k_-  & -m + u_2 k^2    \\
    \end{pmatrix},
    \label{eq:hamiltonian_c6_2}
\end{equation}
where $\beta_1$ is a complex constant and $u_1$ and $u_2$ are real constants.
From the eigenvectors from the effective Hamiltonian~\eqref{eq:hamiltonian_c6_2}, by setting $m \to 0$, we obtain in the limit of $k \to 0$
\begin{equation}
    \det V_{\bm{k}} = +1
    \label{eq:det_V_k_c6}
\end{equation}
for any value of $\phi$, which means it in Case (A).
Thus, when the band gap of a $C_6$-symmetric system closes at TRIM between the pairs, (c)$(\bar{\Gamma}_{10} \bar{\Gamma}_{11}, \bar{\Gamma}_7 \bar{\Gamma}_ 8)$, a $Z_2$ topological phase transition does not occur.

\subsection{\texorpdfstring{$S_4$}{S4} symmetry}\label{sec:s4}
At the TRIM under the $S_4$ symmetry and time-reversal symmetry, there are two irreps forming Kramers doublets: $\bar{\Gamma}_6 \bar{\Gamma}_8$ with $S_4 = e^{\pm \pi i / 4}$ and $\bar{\Gamma}_5 \bar{\Gamma}_7$ with $S_4 = e^{\pm 3 \pi i / 4}$.
Thus, we need to consider a band inversion between the irreps $\bar{\Gamma}_6 \bar{\Gamma}_8$ and $\bar{\Gamma}_5 \bar{\Gamma}_7$.
The effective Hamiltonian in the vicinity of TRIM for this system is the same as Eq.~\eqref{eq:hamiltonian_c4} with a replacement $k_y \to -k_y$ (or $k_{\pm} \to k_{\mp}$).
Because the sign of $\det V_{\bm{k}}$ is not affected by flipping the sign of $k_y$, we can derive
\begin{equation}
    \det V_{\bm{k}} = -1
    \label{eq:det_V_k_s4}
\end{equation}
in the limit $m \to 0$ and $\bm{k} \to 0$.
Therefore, if the system has $S_4$ symmetry, the $Z_2$ topological phase transition always occurs when the band gap closes at TRIM.

\subsection{\texorpdfstring{$C_{3h}$}{C3h} symmetry}\label{sec:c3h}
At the TRIM under the $C_{3h}$ symmetry and time-reversal symmetry, there are three irreps, $\bar{\Gamma}_{10} \bar{\Gamma}_{11}$ with $I C_6 = e^{\pm \pi i / 6}$, $\bar{\Gamma}_7 \bar{\Gamma}_8$ with $I C_6 = e^{\pm  \pi i / 2}$ and $\bar{\Gamma}_9 \bar{\Gamma}_{12}$ with $I C_6 = e^{\pm 5 \pi i / 6}$, all of which are doubly degenerate due to time-reversal symmetry, where $I$ represents the space inversion.
Thus, we need to consider three cases for band inversions between the pairs of irreps (a)$(\bar{\Gamma}_{10} \bar{\Gamma}_{11}, \bar{\Gamma}_7 \bar{\Gamma}_8)$, (b)$(\bar{\Gamma}_7 \bar{\Gamma}_8, \bar{\Gamma}_9 \bar{\Gamma}_{12})$ and (c)$(\bar{\Gamma}_9 \bar{\Gamma}_{12}, \bar{\Gamma}_{10} \bar{\Gamma}_{11})$.

First, the effective Hamiltonian in the two cases (a)$(\bar{\Gamma}_{10} \bar{\Gamma}_{11}, \bar{\Gamma}_7 \bar{\Gamma}_8)$ and (b)$(\bar{\Gamma}_7 \bar{\Gamma}_{8}, \bar{\Gamma}_{9} \bar{\Gamma}_{12})$ is
\begin{equation}
    H(\bm{k}) =
    \begin{pmatrix}
        m                & 0              & 0                & \alpha_1 k_{+} \\
        0                & m              & \alpha_1^* k_{-} & 0              \\
        0                & \alpha_1 k_{+} & -m               & 0              \\
        \alpha_1^* k_{-} & 0              & 0                & -m
    \end{pmatrix},
    \label{eq:hamiltonian_c3h_11}
\end{equation}
where $\alpha_1$ is a complex parameter.
From the eigenvectors from the effective Hamiltonian~\eqref{eq:hamiltonian_c3h_11}, by setting $m \to 0$, we obtain
\begin{equation}
    \det V_{\bm{k}} = -1
    \label{eq:det_V_k_c3h_1}
\end{equation}
and the $Z_2$ topological invariant changes.

Second, the effective Hamiltonian in the case (c)$(\bar{\Gamma}_9 \bar{\Gamma}_{12}, \bar{\Gamma}_{10} \bar{\Gamma}_{11})$ is
\begin{equation}
    H(\bm{k}) =
    \begin{pmatrix}
        m              & 0              & \alpha_1 k_- & 0                \\
        0              & m              & 0            & - \alpha_1^* k_+ \\
        \alpha_1^* k_+ & 0              & -m           & 0                \\
        0              & - \alpha_1 k_- & 0            & -m
    \end{pmatrix},
    \label{eq:hamiltonian_c3h_21}
\end{equation}
where $\alpha_1$ is a complex parameter, and we obtain
\begin{equation}
    \det V_{\bm{k}} = -1,
    \label{eq:det_V_k_c3h_2}
\end{equation}
and the $Z_2$ topological invariant changes.
Thus, if the system has $C_{3h}$ symmetry, the $Z_2$ topological phase transition occurs in all the three band inversions (a)--(c) when the band gap closes at TRIM.

We note that since $C_3$ is a subgroup of $C_{3h}$, the results for $C_3$ and $C_{3h}$ should be consistent.
Their irreps are related as
\begin{equation}
    \begin{split}
        \bar{\Gamma}_7 \bar{\Gamma}_8 \, (C_{3h}) &\to \bar{\Gamma}_4 \bar{\Gamma}_4 \, (C_3), \\
        \bar{\Gamma}_9 \bar{\Gamma}_{12} \, (C_{3h}) &\to \bar{\Gamma}_5 \bar{\Gamma}_6 \, (C_3), \\
        \bar{\Gamma}_{10} \bar{\Gamma}_{11} \, (C_{3h}) &\to \bar{\Gamma}_5 \bar{\Gamma}_6 \, (C_3).
    \end{split}
\end{equation}
We can directly see that the cases (a) $(\bar{\Gamma}_{10} \bar{\Gamma}_{11}, \bar{\Gamma}_7 \bar{\Gamma}_8)$ and (b) $(\bar{\Gamma}_7 \bar{\Gamma}_8, \bar{\Gamma}_9 \bar{\Gamma}_{12})$ in $C_{3h}$ are consistent with the result in $C_3$.
Meanwhile, in the case (c) $(\bar{\Gamma}_9 \bar{\Gamma}_{12}, \bar{\Gamma}_{10} \bar{\Gamma}_{11})$, the occupied and unoccupied states at the TRIM both follow the same irreps $\bar{\Gamma}_5 \bar{\Gamma}_6$ in $C_3$, and the gap cannot close under $C_3$.
At first sight, it may look like a contradiction, but through a detailed analysis in Appendix~\ref{sec:C3h-C3}, we can show that they are consistent.
In short, when a small $C_3$-breaking term is introduced, the gap closing deviates from $\bm{k} = 0, m = 0$ to $\bm{k} \neq 0$ and $m \neq 0$, which does not contradict the fact that the gap does not close at $\bm{k} = 0, m = 0$ in $C_3$ in the case (c).

\begin{figure}[t]
    \centering
    \includegraphics[width = 8cm]{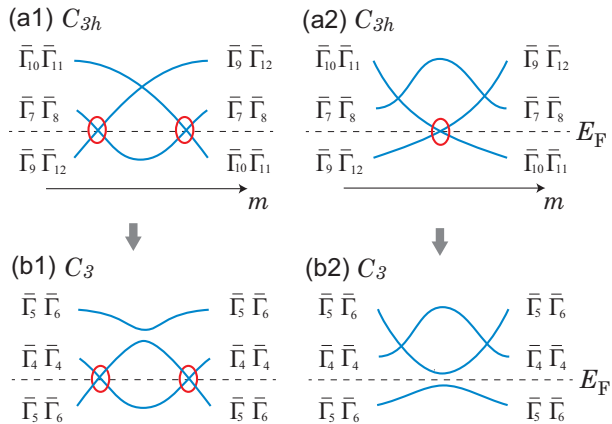}
    \caption{Multiple band inversions at $\Gamma$ in $C_{3h}$ symmetric systems via the change of the parameter $m$. (a1) and (a2) are two examples of multiple band inversions. While they are considered to be adiabatically connected, the change of the $Z_2$ topological invariant $\nu$ is 0 in (a1) and 1 in (a2). This apparent contradiction is resolved by considering that there should be gap-closing points away from $\Gamma$ (not shown in the figure) in (a2) (or (a1)), which gives an additional change of the $Z_2$ topological invariant $\nu$. This point can also be confirmed by lowering the symmetry from $C_{3h}$ to $C_3$. 	Under this symmetry lowering, the change of levels in (a1) and (a2) will become (b1) and (b2), respectively,
        Both in (b1) and (b2), the total change of the $Z_2$ topological invariant is zero, and there is no contradiction. The change from (a2) to (b2) is caused by the coalescence of the off-$\Gamma$ gap-closing points.}
    \label{fig:C3h-Gamma}
\end{figure}

It may look contradictory that all the three band inversions (a)--(c) lead to a $Z_2$ topological phase transition, but in fact, it does not lead to a contradiction, as we show in the following. Let us consider the changes of the levels at $\Gamma$ as we change the parameter $m$, as shown in Figs.\ref{fig:C3h-Gamma} (a1) and (a2). These two cases seem to be adiabatically changed to each other, while from the result in this subsection, (a2) leads to a $Z_2$ topological phase transition, while (a1) does not, because the number of band inversions (red circles in the figure) is different between these two cases.

This problem is resolved as follows. There should be a band-gap closing away from the $\Gamma$ point in (a2) (or (a1)), which also gives rise to the change of the $Z_2$ topological invariant.
Such gap-closing points away from $\Gamma$ are not shown in Fig.~\ref{fig:C3h-Gamma}, where only the levels at $\Gamma$ are shown.
To see the existence of such gap-closing points in a simple way, suppose we add a perturbation to lower the symmetry from $C_{3h}$ to $C_3$.
Under this symmetry lowering, the change of levels in Figs.~\ref{fig:C3h-Gamma} (a1) and (a2) will become (b1) and (b2),  respectively,
because $\bar{\Gamma}_{10}\bar{\Gamma}_{11}$ and $\bar{\Gamma}_{9}\bar{\Gamma}_{12}$ become identical under $C_3$, making the band crossings between them to anticrossings.
Both in (b1) and (b2) the total change of the $Z_2$ topological invariant is zero, and there is no contradiction.
From (a2) to (b2), as discussed in Appendix \ref{sec:C3h-C3},  by the lowering of symmetry, gap-closing points away from the $\Gamma$ point will coalesce to the gap closing at the $\Gamma$
point. It shows the existence of gap-closing points away from $\Gamma$ in (a2). Thus, we conclude that there should be a gap closing points away from $\Gamma$ in (a2), which is consistent with (a1) on the unchanged $Z_2$ topological invariant.

\subsection{Summary of Sec.~\ref{sec:lg}}
\begin{table}[tbp]
    \centering
    \begin{tabular}{cccc}
        PGs      & irreps                                                                      & $\det V_k$ & phase transition \\ \hline
        $C_3$    & $(\bar{\Gamma}_4 \bar{\Gamma}_4, \bar{\Gamma}_5 \bar{\Gamma}_6)$            & $-1$       & \Checkmark       \\
        $C_4$    & $(\bar{\Gamma}_6 \bar{\Gamma}_8, \bar{\Gamma}_5 \bar{\Gamma}_7)$            & $-1$       & \Checkmark       \\
        $C_6$    & $(\bar{\Gamma}_{10} \bar{\Gamma}_{11}, \bar{\Gamma}_7 \bar{\Gamma}_8)$      & $-1$       & \Checkmark       \\
        $C_6$    & $(\bar{\Gamma}_{9} \bar{\Gamma}_{12}, \bar{\Gamma}_{10} \bar{\Gamma}_{11})$ & $+1$       & \XSolidBrush     \\
        $C_6$    & $(\bar{\Gamma}_{7} \bar{\Gamma}_{8}, \bar{\Gamma}_9 \bar{\Gamma}_{12})$     & $-1$       & \Checkmark       \\
        $S_4$    & $(\bar{\Gamma}_{6} \bar{\Gamma}_{8}, \bar{\Gamma}_5 \bar{\Gamma}_{7})$      & $-1$       & \Checkmark       \\
        $C_{3h}$ & $(\bar{\Gamma}_{10} \bar{\Gamma}_{11}, \bar{\Gamma}_{7} \bar{\Gamma}_{8})$  & $-1$       & \Checkmark       \\
        $C_{3h}$ & $(\bar{\Gamma}_{9} \bar{\Gamma}_{12}, \bar{\Gamma}_{10} \bar{\Gamma}_{11})$ & $-1$       & \Checkmark       \\
        $C_{3h}$ & $(\bar{\Gamma}_{7} \bar{\Gamma}_{8}, \bar{\Gamma}_{9} \bar{\Gamma}_{12})$   & $-1$       & \Checkmark       \\\hline
    \end{tabular}
    \caption{Five point groups (PGs) ($C_3, C_4, C_6, S_4,$ and $C_{3h}$) where $Z_2$ topological phase transition occurs at high-symmetry points. For each PG, two occupied and two unoccupied bands near the Fermi energy can take irreducible representations (irreps) in the table. For all these irreps, we obtain $\det V_{\bm{k}}$ to determine whether the $Z_2$ topological phase transition occurs. The $Z_2$ topological phase transition does not occur if and only if the pair of irreps of the point group $C_6$ are $(\bar{\Gamma}_{9} \bar{\Gamma}_{12}, \bar{\Gamma}_{10} \bar{\Gamma}_{11})$.}
    \label{tb:PGs}
\end{table}

In Sec.~\ref{sec:lg}, we have explained our theory of the 2D $Z_2$ topological phase transition under the general layer group symmetry.
Table~\ref{tb:PGs} presents the results obtained from the discussion in Sec.~\ref{sec:c3} and Sec.~\ref{sec:4}--\ref{sec:c3h}.
The values of $\det V_{\bm{k}}$ at $\bm{k} \to 0, m \to 0$ corresponding to each point group and irreps at TRIM are shown in Table~\ref{tb:PGs}.
As we discussed, $\det V_{\bm{k}} = +1$ and $-1$ mean the cases (A) and (B), corresponding to the absence and presence of a $Z_2$ transition, respectively.
Thus, we find that among the cases considered, the $Z_2$ topological phase transition does not occur if and only in the pair of irreps of the point group $C_6$ are $(\bar{\Gamma}_{9} \bar{\Gamma}_{12}, \bar{\Gamma}_{10} \bar{\Gamma}_{11})$.

\section{discussion}\label{sec:discussion}
This paper is inspired by the previous work on discontinuous changes of the piezoelectric tensor associated with 2D $Z_2$ topological phase transitions~\cite{Yu2020-ik}.
It exhausts gap-closing at all points in $\bm{k}$-space, including HSPs, for all the plane groups that do not have inversion or two-fold rotation symmetry.
The work in Ref.~\cite{Yu2020-ik} is regarded as an extension of Refs.~\cite{Murakami2007-gu,Murakami2007-lt} on 2D $Z_2$ topological phase transitions into general point group symmetries.
We note the following two points in the comparison between our paper and Ref.~\cite{Yu2020-ik}.
First, Ref.~\cite{Yu2020-ik} studies the 7 plane groups without inversion or $C_2$ symmetries among the 17 plane groups.
Meanwhile, our paper discusses the 80 layer groups, which include 17 plane groups.
Second, our analysis is complementary to the approach in Ref.~\cite{Yu2020-ik} in some respects. In Ref.~\cite{Yu2020-ik}, in particular in its Supplementary Note 2.C.1, it is shown that a band inversion between different $C_3$ doublets
($\bar{\Gamma}_4\bar{\Gamma}_4$ and $\bar{\Gamma}_5\bar{\Gamma}_6$ in our notation) in
$C_3$-symmetric systems is always accompanied by a change of
the $Z_2$ topological invariant $\nu$. In this respect, this result is in perfect agreement with our result in Sec.~\ref{sec:c3} in the present paper. Meanwhile, the approach in Ref.~\cite{Yu2020-ik} toward this conclusion is different from ours, and is basically based on the continuity argument. Namely, if we gradually change system parameters so that the $C_3$-symmetric system acquires inversion symmetry, then the band inversion is shown to be accompanied by a $Z_2$ phase transition. Then it is discussed in Ref.~\cite{Yu2020-ik} that the change of the $Z_2$ transition at the band inversion will stay the same even when the inversion symmetry is broken. This argument is correct, but it relies on the assumption that the system can be adiabatically connected to an inversion-symmetric system with keeping the $C_3$ symmetry. This assumption is obvious for simple models, but for complex systems and real materials, it is not necessarily obvious.
In this sense, our argument in Sec.~\ref{sec:c3} gives proof for the change of the $Z_2$ topological invariant in $C_3$ symmetric systems without relying on such assumptions, and is complementary to the theory in Ref.~\cite{Yu2020-ik}.
Thus, in this paper, we construct all the topological phase transitions in all the layer groups, which is an extension of the plane groups, including symmetry operations flipping the two surfaces of the 2D system.
Therefore, together with the works in Refs.~\cite{FuKane-inversion,Murakami2007-gu,Murakami2007-lt,Yu2020-ik}, we have covered all the $Z_2$ topological phase transitions in 2D systems.

Our results are new and not included in the theory of symmetry-based indicators~\cite{Po2017-mk,Khalaf-PRX} and topological quantum chemistry~\cite{Bradlyn2017-ot,Bradlyn2018-eu}.
In systems with some crystallographic symmetry, topological invariants are written only in terms of irreps at HSPs, as is well understood in terms of topological quantum chemistry~\cite{Bradlyn2017-ot,Bradlyn2018-eu} and symmetry-based indicators~\cite{Po2017-mk,Khalaf-PRX}.
The topological phase transitions and band inversions are directly related in such cases.
Meanwhile, the $Z_2$ topological invariant $\nu$ in 2D systems becomes a symmetry-based indicator if and only if the inversion symmetry is present.
Namely, it is limited to the layer groups with ``inv" in Table~\ref{tb:80LGs}.
Therefore, $Z_2$ topological phase transitions for the other layer groups are outside the topological quantum chemistry and symmetry-based indicator theory.

\section{conclusion}\label{sec:conclusion}
This paper describes all 2D $Z_2$ topological phase transition patterns by band inversions.
The behavior of gap-closing at high-symmetry points has not been well understood in systems in the absence of inversion symmetry.
We examined the behavior of gap-closing at TRIM for all of the 80 layer groups describing 2D systems.
The results are shown in Table~\ref{tb:80LGs}.
We found that the layer groups with a blank entry in Table~\ref{tb:80LGs} can have nontrivial topological phase transitions at the band inversion.
The point group symmetry can describe the band inversion at the high-symmetry point where the gap closes.
The correspondence between the layer groups and the point groups at the high-symmetry points considered is shown in Table~\ref{tb:17LGs}.
The point groups to consider are limited to the subgroups of the point groups shown in Table~\ref{tb:17LGs}, which are $C_3, C_4, C_6, S_4,$ and $C_{3h}$.
Whether the $Z_2$ topological phase transition occurs is determined only from the irreps of two occupied bands and two unoccupied bands near the Fermi energy, which are calculated in detail as shown in Table~\ref{tb:PGs}.
In particular, the results for $C_3$ are described in detail in Sec.~\ref{sec:c3}.
In addition to the gap closing at TRIM, here we mention the gap closing at not-TRIM, including non-TRIM high-symmetry points such as $K$ and $K'$ points.
At the gap closing at non-TRIM, the change of the $Z_2$ topological invariant is equal to half of the number of gap closing points in the Brillouin zone, which follows from Refs.~\cite{Murakami2007-gu,Murakami2007-lt}.

These results show that whether or not the $Z_2$ topological phase transition occurs at the band inversion is entirely determined by the irreps involved in band inversion.
It is a nontrivial result since our conclusion also applies to all the cases where the $Z_2$ topological invariant is not described in terms of symmetry-based indicator or topological quantum chemistry.

\begin{acknowledgments}
    We thank Tiantian Zhang for the fruitful discussions.
    This work was partly supported by JSPS KAKENHI Grants No. JP20H04633, JP21K13865 and JP22H00108.
\end{acknowledgments}

\appendix

\section{Detailed calculations on a \texorpdfstring{$C_3$-symmetric}{C3-symmetric} system}\label{sec:detailed}
In the main text, we explained that by defining $v_\alpha(\bm{k}, m)$ by Eq.~\eqref{eq:u_alpha_plus_delta}, with a proper choice of $v_\alpha(\bm{k}, \alpha)$ within $k_1 \leq |\bm{k}| \leq k_2$, we can analytically calculate the change of the $Z_2$ topological invariant $\nu$ across $m = 0$.
In this Appendix, we show details of this calculation.
As we noted earlier, when Eq.~\eqref{eq:gauge} is satisfied, $\det [w]$ and $\mathrm{Pf} [w]$ becomes trivial.
In order to use this fact, we have chosen the gauges of $u_\alpha^{(\pm)} (\bm{k})$ so that they satisfy Eq.~\eqref{eq:gauge} in the vicinity of $(0, 0)$ ($\bm{k} = 0$), including the disk $|\bm{k}| \leq k_2$.
Such a gauge choice only near $(0, 0)$ is always possible.
Namely, we impose
\begin{equation}
    \begin{split}
        \ket{u_1^{(\pm)}(-\bm{k}, m)}  &= \Theta \ket{u_2^{(\pm)}(\bm{k}, m)}, \\
        \ket{u_2^{(\pm)}(-\bm{k}, m)}  &= - \Theta \ket{u_1^{(\pm)}(\bm{k}, m)}
    \end{split}
    \label{eq:rel_in_region_V}
\end{equation}
near $(0, 0)$.
When the wavefunctions $v_\alpha(\bm{k}, m)$ are either $u^{(+)}$ or $u^{(-)}$, the matrix $w$ defined in terms of $v_\alpha$ satisfies $\mathrm{Pf}[w] = 1 = \det[w]$, and they do not contribute to the $Z_2$ topological invariant $\nu$.
Thus, from Eq.~\eqref{eq:u_alpha_plus_delta}, we need to consider only the region $k_1 < |\bm{k}| < k_2$.
From Eq.~\eqref{eq:z2_invariant}, we get
\begin{align}
      & \nu(m = +\delta) - \nu(m = -\delta) \notag                                   \\
    = & - P_\theta^{k_y = 0} (m = +\delta) + P_\theta^{k_y = 0} (m = -\delta) \notag \\
    \begin{split}
        = & - \frac{1}{2 \pi i} \int_{k_1}^{k_2} \dd{k_x} \dv{k_x} \left( \log \det w(k_x, 0)_{m = +\delta} \right. \\
        & \left. \hspace{9em} - \log \det w(k_x, 0)_{m = -\delta} \right),
    \end{split}
    \label{eq:diff_z2_invariant}
\end{align}
where the matrix $w$ is defined in terms of $v_\alpha(\bm{k}, m)$.
When $m = -\delta$, the wavefunctions $v_\alpha (\bm{k}, m)$ are equal to $u_\alpha^{(-)} (\bm{k}, m)$, but when $m = +\delta$, they deviate from $u_\alpha^{(-)}(\bm{k}, m)$ in general, and this difference contributes to Eq.~\eqref{eq:diff_z2_invariant}.
To calculate this difference, we define a unitary matrix $U_{\bm{k}}$ as:
\begin{equation}
    (v_1(\bm{k}, +\delta), v_2(\bm{k}, +\delta)) = (u^{(-)}_1(\bm{k}, +\delta), u^{(-)}_2(\bm{k}, +\delta)) U_{\bm{k}}.
\end{equation}
Using this unitary matrix, we obtain
\begin{equation}
    w(\bm{k})_{m = +\delta} = U_{-\bm{k}}^\dagger w(\bm{k})_{m = -\delta} U_{\bm{k}}^*,
\end{equation}
in the limit $\delta \to +0$, which leads to
\begin{equation}
    \begin{split}
        &\nu(m = +\delta) - \nu(m = -\delta) \\
        = & \frac{1}{2 \pi i} \int_{k_1}^{k_2} \dd{k_x} \dv{k_x}  \left. \left(\log \det U_{-\bm{k}} - \log \det U_{\bm{k}}^* \right) \right|_{\bm{k} = (k_x, 0)}.
    \end{split}
    \label{eq:diff_z2_invariant_2}
\end{equation}

\begin{figure}[t]
    \centering
    \includegraphics[width = 1\columnwidth]{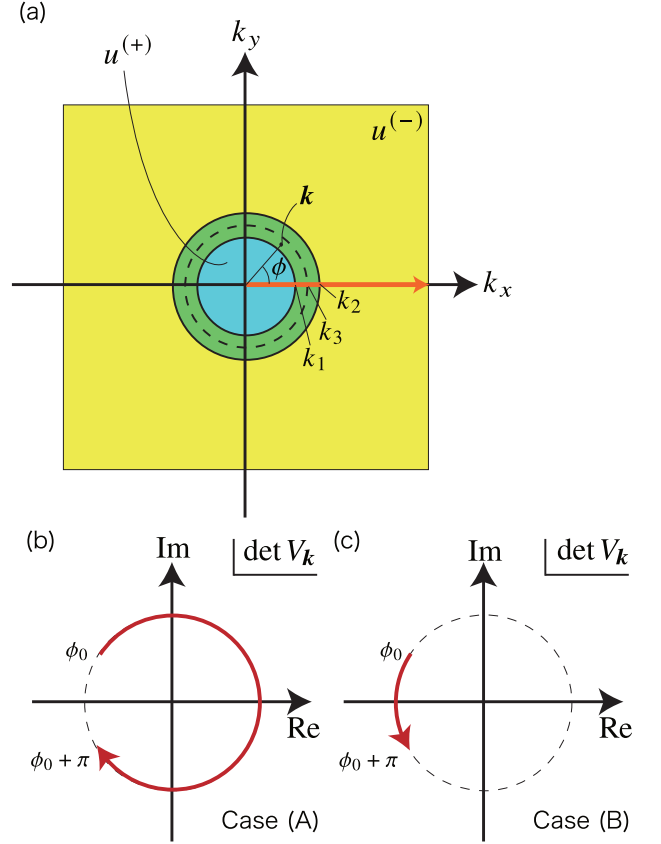}
    \caption{Calculation of the change of the $Z_2$ topological invariant at the band inversion. (a) Eigenstates for $m > 0$. For $k < k_1$, we take $u^{(+)}$, and for $k > k_2$, we take $u^{(-)}$. In the region of $k_1 < k < k_2$, we take the eigenstates, which are smoothly connected from $u^{(+)}$ to $u^{(-)}$. We take $k_3$ such that $k_1 < k_3 < k_2$ and make the Eq.~\eqref{eq:rel_U_k} be satisfied for $k_1 < k < k_3$. (b--c) Whether a $Z_2$ topological phase transition occurs or not depends on whether $\det V_{\bm{k}}$ passes through (b) the positive side of the real axis (Case (A)) or (c) the negative side (Case (B)) when the azimuth angle of $\bm{k}$ is changed from $\phi_0$ to $\phi_0 + \pi$. In Case (B), the $Z_2$ topological phase transition occurs while not in Case (A).}
    \label{fig:Z2-2a_Appendix}
\end{figure}

Thus, our remaining task is to determine $U_{\bm{k}}$ in the region $k_1 \leq k \leq k_2$.
For $k = k_2$, $U_{\bm{k}}$ is the identity matrix $I_2$, and for $k = k_1$, $U_{\bm{k}}$ is equal to $V_{\bm{k}}$ defined by
\begin{equation}
    (u_1^{(+)}(\bm{k}, +\delta), u_2^{(+)}(\bm{k}, +\delta)) = (u^{(-)}_1(\bm{k}, +\delta), u^{(-)}_2(\bm{k}, +\delta)) V_{\bm{k}}.
    \label{eq:def_V_k_appendix}
\end{equation}
From our choice of gauge for $u_i^{(\pm)} (\bm{k}, m)$, it follows that
\begin{equation}
    V_{-\bm{k}} = \sigma_y V^*_{\bm{k}} \sigma_y
    \label{eq:rel_V_k}.
\end{equation}
Now we construct $v_\alpha (\bm{k}, m)$ for $k_1 < k < k_2$.
Namely, we need to construct a unitary matrix $U_{\bm{k}}$ for $k_1 < k < k_2$ such that $U_{\bm{k}}$ continuously interpolates between $U_{\bm{k}} = I_2 \ (|\bm{k}| = k_2)$ and $U_{\bm{k}} = V_{\bm{k}} \ (|\bm{k}| = k_1)$.
When
\begin{equation}
    U_{-\bm{k}} = \sigma_y U_{\bm{k}}^{*} \sigma_y
    \label{eq:rel_U_k}
\end{equation}
is satisfied, the integrand in Eq.~\eqref{eq:diff_z2_invariant_2} becomes zero.
Therefore, we intend to construct the unitary matrix $U_{\bm{k}}$ for $k_1 < k < k_2$ to satisfy Eq.~\eqref{eq:rel_U_k} whenever possible.
For this purpose, we need to study the properties of $V_{\bm{k}}$.
Since $V_{\bm{k}}$ is a unitary matrix, the absolute value of $\det V_k$ is 1, and let $\theta_{\bm{k}} = \arg \det V_{\bm{k}}$ denote its argument.
When we restrict $\bm{k}$ to a circle of radius $k_1$ and regard $\bm{k}$ as a function of its azimuth angle $\phi$, \textit{i}.\textit{e}., $\bm{k} = (k_1 \cos \phi, k_1 \sin \phi)$, we obtain $-\theta_{\phi + \pi} = \theta_{\phi}$ from Eq.~\eqref{eq:rel_U_k}.
Therefore, when $\phi$ is changed along this circle from $\phi_0$ to $\phi_0 + \pi$ for some $\phi_0$,  $\det V_{\bm{k}}$ moves along the unit circle to a position symmetric about the real axis.
This movement is classified into two cases (A) and (B): across the positive part of the real axis or the negative part of the real axis, as shown in Figs.~\ref{fig:Z2-2a_Appendix}(b) and (c), respectively.
More generally, from $\phi_0$ to $\phi_0 + \pi$, $\det V_{\bm{k}}$ crosses the positive (negative) part of the real axis at odd times in Case (A) (Case (B)).
Representative examples of $V_{\bm{k}}$ for the cases (A) and (B) are given by the following matrices:
\begin{equation}
    \begin{aligned}
        \mbox{(A)} \quad & V_{\bm{k}} = I_2                        & \to         & \det V_{\bm{k}} = 1, \\
        \mbox{(B)} \quad & V_{\bm{k}} = Q_{\bm{k}} \equiv i \mqty( & e^{-i \phi}                        \\ -e^{i\phi} & ) & \to & \det V_{\bm{k}} = -1.
    \end{aligned}
    \label{eq:V_to_I_2}
\end{equation}

To fix $U_{\bm{k}}$ in the region $k_1 < k < k_2$, we divide this region by a circle with a radius $k_3$ with $k_1 < k_3 < k_2$, as shown in Fig.~\ref{fig:Z2-2a_Appendix}(a).
To simplify the calculation, we will determine $U_{\bm{k}}$ within the inner region $k_1 < k < k_3$ with keeping the condition~\eqref{eq:rel_U_k} so that the matrix $U_{\bm{k}}$ on the circle $k = k_3$ is equal to $I_2$ or $Q_{\bm{k}}$, for the cases (A) or (B), respectively.
To this end, we separate the unitary matrix $U_{\bm{k}}$ into the determinant $\det U_{\bm{k}}$ and a unitary matrix $M_{\bm{k}}$ with a determinant equal to unity:
\begin{equation}
    U_{\bm{k}} = \sqrt{\det U_{\bm{k}}} M_{\bm{k}}, \quad \det M_{\bm{k}} = 1.
    \label{eq:Uk_and_detMk}
\end{equation}
From $k = k_1$ to $k = k_3$, we gradually change $\det U_{\bm{k}}$ so that at $k = k_3$, $\det U_{\bm{k}}$ is constant equal to $1$ or $-1$ for the cases (A) or (B), respectively.
Here, the branch of the square root in Eq.~\eqref{eq:Uk_and_detMk} is taken to be continuous in $\bm{k}$, and at $k = k_3$, we take the branch $\sqrt{-1} = i$ in the case (B).
On the other hand $M_{\bm{k}} \in \mathrm{SU}(2)$ is expressed as
\begin{equation}
    M_{\bm{k}} = \begin{pmatrix}
        \alpha_0 + i \alpha_3   & \alpha_1 - i \alpha_2 \\
        - \alpha_1 - i \alpha_2 & \alpha_0 - i \alpha_3
    \end{pmatrix}, \quad \alpha_i \in \mathbb{R},  \quad \sum_{i = 0}^{3} \alpha_i^2 = 1,
\end{equation}
and therefore there is one-to-one correspondence between $M_{\bm{k}}$ and a point $P_{\bm{k}}(\alpha_0, \alpha_1, \alpha_2, \alpha_3)$ on a three-dimensional unit sphere $S^3: \sum_{i = 0}^3 x_i^2 = 1$.
As $\bm{k}$ is changed along the circle $k = k_1$, the point $P_{\bm{k}}$ draws a loop on $S^3$.
In the case (A), from $k = k_1$ to $k = k_3$, we set $M_{\bm{k}}$ so that the corresponding loop continuously shrinks to a point $(1, 0, 0, 0)$, meaning that $M_{\bm{k}} = I_2$ on the circle $k = k_3$.
Here we impose a condition in Eq.~\eqref{eq:rel_U_k} throughout this change from $k = k_1$ to $k = k_3$.
This means that $P_{\bm{k}} = P_{-\bm{k}}$, and one can construct $M_{\bm{k}}$ so that this condition is always satisfied within $k_1 \leq k \leq k_3$.
On the other hand, in the case (B), we impose Eq.~\eqref{eq:rel_U_k} within $k_1 \leq k \leq k_3$.
Then it follows that $M_{\bm{k}}$ satisfies $M_{-\bm{k}} = -\sigma_y M_{\bm{k}}^* \sigma_y$ because of the branch choice of $\sqrt{\det U_{\bm{k}}}$.
It means we impose $-P_{\bm{k}} = P_{-\bm{k}}$ within $k_1 \leq k \leq k_3$ in the case (B).
By considering this constraint, we construct $M_{\bm{k}}$ so that it is equal to $\mqty( & e^{-i \phi}\\ -e^{i\phi} & )$ at $k = k_3$, which means $P_{\bm{k}}$ is given by $(0, \cos\phi, \sin\phi, 0)$ at $k = k_3$.
By this method, we can construct $U_{\bm{k}}$ within $k_1 \leq k \leq k_3$ to satisfy Eq.~\eqref{eq:rel_U_k}.

Next, we define $U_{\bm{k}}$ for $k_3 < k < k_2$ to be $U_{\bm{k}} = I_2$ for the case (A) (Fig.~\ref{fig:Z2-2a_Appendix}(b)) and $U_{\bm{k}} = e^{\frac{\pi i}{2} (1 - t)} \mqty(t & \sqrt{1 - t^2} e^{-i \phi}\\ - \sqrt{1 - t^2} e^{i\phi} & t)$ for the case (B) (Fig.~\ref{fig:Z2-2a_Appendix}(c)), where $t$ is a parameter such that $t = 0$ when $k = k_3$ and $t = 1$ when $k = k_2$.

Now we calculate Eq.~\eqref{eq:diff_z2_invariant_2}.
Because the matrix $U_{\bm{k}}$ in the region $k_1 < k < k_3$ satisfies Eq.~\eqref{eq:rel_U_k}, it does not contribute to Eq.~\eqref{eq:diff_z2_invariant_2}.
Thus, only the region between $k = k_2$ and $k = k_3$ contributes to Eq.~\eqref{eq:diff_z2_invariant_2}, and we get
\begin{equation}
    \nu(m = +\delta) - \nu(m = -\delta) = 0
\end{equation}
in the case (A) and
\begin{equation}
    \begin{split}
        &\nu(m = +\delta) - \nu(m = -\delta)\\
        = &\frac{1}{2 \pi i} \int_0^1 \dd{t} \dv{t} [2 \pi i (1 - t)] = -1
    \end{split}
\end{equation}
in the case (B).

Therefore, we can summarize these results to say that the difference in $Z_2$ topological invariants is determined by how $\det V_{\bm{k}}$ moves on the unit circle when the argument of $\bm{k}$ is changed from $\phi_0$ to $\phi_0 + \pi$, and we conclude
\begin{equation}
    \nu(m = +\delta) - \nu(m = -\delta) =
    \left\{
    \begin{aligned}
         & 0 \qq{: case (A)}  \\
         & 1 \qq{: case (B)}.
    \end{aligned}
    \right.
    \label{eq:nu-nu_case_appendix}
\end{equation}

\section{Tight-binding model with \texorpdfstring{$C_3$}{C3} symmetry}\label{sec:TBmodel}
\begin{figure}[tb]
    \includegraphics[width=1\columnwidth]{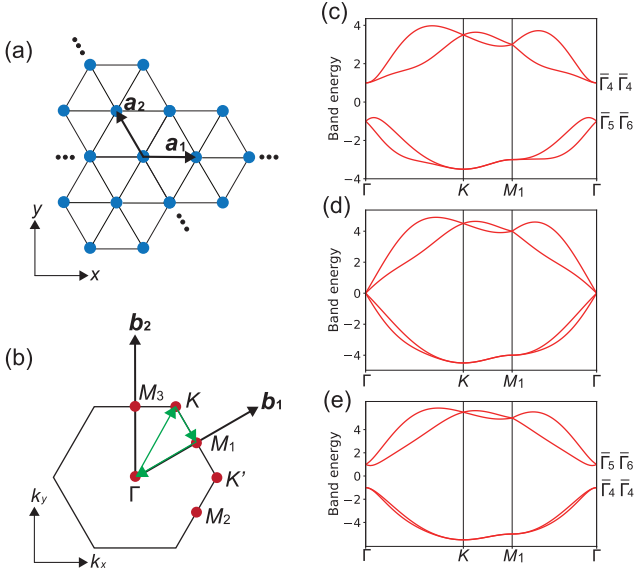}
    \caption{The tight-binding model with $C_{3}$ symmetry. (a) A simple triangular lattice of the model. (b) The Brillouin zone and the high-symmetry points for this model. (c--e) The bulk band structures of this model with (c) $m'=2$,   (d) $m'=3$, and (e) $m'=4$. }\label{fig:TB_model}
\end{figure}

In this section, to confirm our theory of the $Z_2$ topological phase transitions in Sec.~\ref{sec:c3}, we introduce a 2D tight-binding model with time-reversal and $C_3$ symmetry.
This model is defined on a simple triangular lattice, where the lattice vectors are given by $\boldsymbol{a}_1=(1,0)$ and $\boldsymbol{a}_2=(-1/2, \sqrt{3}/2)$ [Fig.~\ref{fig:TB_model}(a)].
Each site has two orbitals and spin $\uparrow \downarrow$, and therefore each site has four degrees of freedom.
Our model on this lattice can be expressed as the following four-band Bloch Hamiltonian:
\begin{align}
    H'(\boldsymbol{k})= & (m'-tf_{0} (\boldsymbol{k}))\tau_{z}\otimes \sigma_{0} \nonumber                                                       \\
                        & +\frac{v_{a}}{2}(\tau_{0}+\tau_{z})\otimes (f_{1}(\boldsymbol{k})\sigma_{x}+ f_{2}(\boldsymbol{k})\sigma_{y})\nonumber \\
                        & +v_{b}\tau_{x}\otimes (f_{1}(\boldsymbol{k})\sigma_{x}-f_{2}(\boldsymbol{k})\sigma_{y}),
\end{align}
with
\begin{align}
    f_{0}(\boldsymbol{k}) & =\cos \boldsymbol{k}\cdot \boldsymbol{a}_1 +\cos \boldsymbol{k}\cdot
    \boldsymbol{a}_2 +\cos \boldsymbol{k}\cdot (\boldsymbol{a}_1+\boldsymbol{a}_2), \nonumber                                                                                                                            \\
    f_{1}(\boldsymbol{k}) & =\sin \boldsymbol{k}\cdot \boldsymbol{a}_{1}-\tfrac{1}{2}\sin \boldsymbol{k}\cdot \boldsymbol{a_{2}}+\tfrac{1}{2}\sin \boldsymbol{k}\cdot (\boldsymbol{a}_{1}+\boldsymbol{a}_{2}), \nonumber \\
    f_{2}(\boldsymbol{k}) & =\tfrac{\sqrt{3}}{2}\sin \boldsymbol{k}\cdot \boldsymbol{a}_{2}+\tfrac{\sqrt{3}}{2} \sin \boldsymbol{k} \cdot (\boldsymbol{a}_{1}+\boldsymbol{a}_{2}),
\end{align}
where $\sigma_{0}$ and $\tau_{0}$ are $2\times 2$ identity matrices, $\tau_{i}$ $(i=x,y,z)$ are the Pauli matrices corresponding to the two orbitals at each site, and $\sigma_{i}$ $(i=x,y,z)$ are the Pauli matrices corresponding to the spin $\uparrow \downarrow$.
$m^\prime, t, v_a,$ and $v_b$ are real parameters.

Next, we discuss symmetries in this model. This model has time-reversal symmetry: $\Theta {H}'(\boldsymbol{k})\Theta^{-1}={H}'(-\boldsymbol{k})$, where $\Theta=-i \tau_{0}\otimes \sigma_{y}K$ with $K$ being  the complex conjugation.  In addition, our model has $C_{3}$ symmetry. The matrix representation of the $C_{3}$ operation is given by
\begin{align}
    C_{3}=\begin{pmatrix}
              \exp(-i\tfrac{\pi}{3}\sigma_{z}) & 0           \\
              0                                & -\sigma_{0} \\
          \end{pmatrix}.
\end{align}
Under $C_{3}$, the wave vector transforms as $\boldsymbol{k}=(k_{x}, k_{y})\rightarrow C_{3}\boldsymbol{k}= (-\tfrac{1}{2}k_{x}-\tfrac{\sqrt{3}}{2}k_{y}, \tfrac{\sqrt{3}}{2}k_{x}-\tfrac{1}{2}k_{y})$, and our model has $C_{3}$ symmetry:
\begin{equation}
    C_{3}H'(\boldsymbol{k})C_{3}^{-1}=H'(C_{3}\boldsymbol{k}).
\end{equation}
Figure \ref{fig:TB_model}(b) shows a Brillouin zone and high-symmetry points of this model, where the reciprocal lattice vectors are given by $\boldsymbol{b}_{1}=2\pi(1, 1/\sqrt{3})$ and $\boldsymbol{b}_{2}=2\pi (0, 2/\sqrt{3})$, and TRIM are $\Gamma=(0,0)$, $M_{1}=\boldsymbol{b}_{1}/2$, $M_{2}=(\boldsymbol{b}_{1}+\boldsymbol{b}_{2})/2$, $M_{3}=\boldsymbol{b}_{2}/2$. Then the $C_{3}$-invariant momenta are given by $K=\boldsymbol{b}_{1}/3+\boldsymbol{b}_{2}/3$ and $K'=2\boldsymbol{b}_{1}/3-\boldsymbol{b}_{2}/3$.

To study topological phase transitions in this model, we calculate band structures of this model with various parameters. Figures \ref{fig:TB_model}(c-e) show the band structures of the bulk Hamiltonian $H'(\boldsymbol{k})$ with the parameters being $t=1$, $v_{a}=0.7$, and $v_{b}=1.3$.  The irreps of $\Gamma$ point for occupied states  are $\bar{\Gamma}_{5}\bar{\Gamma}_{6}$ in Fig.~\ref{fig:TB_model}(c) ($m'=2$) and $\bar{\Gamma}_{4}\bar{\Gamma}_{4}$ in Fig.~\ref{fig:TB_model}(e) ($m'=4$).  This is because the band gap closes at $\Gamma$ point when $m'=3$, resulting in the band inversion.

\begin{figure}[tb]
    \includegraphics[width=1\columnwidth]{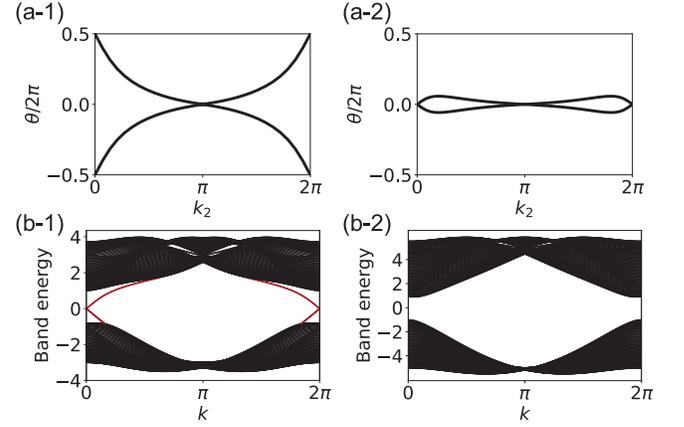}
    \caption{Band structures and Wannier centers for the $C_3$-symmetric model. (a) The evolutions of the Wannier function centers in the model $H'(\boldsymbol{k})$ with (a-1) $m'=2$  ($\nu=1$) and (a-2) $m'=4$ ($\nu=0$). (b) The band structures in the geometry with open boundary conditions in the $\boldsymbol{a}_{1}$ direction with (b-1) $m'=2$  and (b-2) $m'=4$. }\label{fig:TB_edge_WL}
\end{figure}

We compute the $Z_{2}$ topological invariant $\nu$ for $m'=2$ and $m'=4$ in order to see whether the topological phase transition occurs or not.
To compute $\nu$, we introduce the following ${\rm U}(2N)$ matrix \cite{PhysRevB.84.075119,  PhysRevB.89.155114}:
\begin{align}
     & [\mathcal{W}(k_{2})]_{mn} \equiv & \biggl[ \mathcal{P} \exp \Bigl( i\int_{0}^{2\pi} dk_{1} \mathcal{A}_{1} (k_{1},k_2) \Bigr)  \biggr]_{mn},
\end{align}
where $k_{1}$ and $k_{2}$ are the wave vectors in the $\boldsymbol{b}_{1}$ and $\boldsymbol{b}_{2}$ directions, respectively, and $m$ and $n$ run over the band indices of the $2N$ occupied bands. Here $\mathcal{P}$ denotes the path-ordered operation, and $\mathcal{A}_{1} (\boldsymbol{k})$ is a non-Abelian Berry connection
\begin{equation}
    [\mathcal{A}_{1} (\boldsymbol{k}) ]_{mn}=i\bra{u_{\boldsymbol{k},m}}\partial_{k_{1}} \ket{u_{\boldsymbol{k},n}}.
\end{equation}
The matrix $\mathcal{W}(k_{2})$ is a unitary matrix, and therefore the eigenvalues of $\mathcal{W}(k_{2})$
can be expressed as $\exp [i\theta_{n}(k_{2})]$ with $n=1,\cdots , 2N$. The phases $\theta_{n}(k_{2})$ are Wannier function  centers of occupied bands.
By tracking the evolution of the Wannier function centers, we can obtain $\nu$ \cite{PhysRevB.83.235401, PhysRevB.84.075119,  PhysRevB.89.155114}.
Figure \ref{fig:TB_edge_WL}(a) shows the evolutions of $\theta_{n}(k_{2})$ in  our model with $m'=2$ [Fig.~\ref{fig:TB_edge_WL}(a-1)] and $m'=4$ [Fig.~\ref{fig:TB_edge_WL}(a-2)]. In the former case, the spectra of $\theta_{n}(k_{2})$ wind as a function of $k_2$, which means that $\nu=1$. On the other hand, the latter case shows that the spectra do not wind, namely, $\nu=0$.

The nontrivial $Z_{2}$ topological invariant leads to the emergence of edge modes.
Figure~\ref{fig:TB_edge_WL}(b) shows the band structures in the geometry with open boundary conditions in the $\boldsymbol{a}_{1}$ direction. In Fig.~\ref{fig:TB_edge_WL}(b-1) ($m'=2$), $\nu=1$ leads to the gapless edge modes. On the other hand,  in Fig.~\ref{fig:TB_edge_WL}(b-2) ($m'=4$), $\nu=0$ results in the absence of the gapless edge modes.
From these calculations, we conclude that our model with $m'=2$ is a topological insulator and that with $m'=4$ is a trivial insulator, and the gap closing at $\Gamma$ point leads to the $Z_{2}$ topological phase transition.
This confirms the result in Sec.~\ref{sec:c3}.

\section{Relationship between the band inversions under \texorpdfstring{$C_{3h}$ and those under $C_3$}{C3h and those under C3}}\label{sec:C3h-C3}
In this section, we verify consistency in the results for the point groups $C_{3h}$ and $C_3$ in Table~\ref{tb:PGs} of Sec.~\ref{sec:lg}.
Since $C_3$ is a subgroup of $C_{3h}$, the results for $C_3$ and $C_{3h}$ should be consistent.
Their irreps are related as
\begin{equation}
    \begin{split}
        \bar{\Gamma}_7 \bar{\Gamma}_8 \, (C_{3h}) &\to \bar{\Gamma}_4 \bar{\Gamma}_4 \, (C_3), \\
        \bar{\Gamma}_9 \bar{\Gamma}_{12} \, (C_{3h}) &\to \bar{\Gamma}_5 \bar{\Gamma}_6 \, (C_3), \\
        \bar{\Gamma}_{10} \bar{\Gamma}_{11} \, (C_{3h}) &\to \bar{\Gamma}_5 \bar{\Gamma}_6 \, (C_3).
    \end{split}
\end{equation}
Thus, among the three cases in $C_{3h}$ the results for (a) $(\bar{\Gamma}_{10} \bar{\Gamma}_{11}, \bar{\Gamma}_7 \bar{\Gamma}_8)$ and (b) $(\bar{\Gamma}_7 \bar{\Gamma}_8, \bar{\Gamma}_9 \bar{\Gamma}_{12})$ in $C_{3h}$ reduce to $(\bar{\Gamma}_4 \bar{\Gamma}_4, \bar{\Gamma}_5 \bar{\Gamma}_6)$ in $C_3$ studied in Sec.~\ref{sec:c3}, and they are consistent in that they lead to $Z_2$ topological phase transitions.
Meanwhile, in the case (c) $(\bar{\Gamma}_{10} \bar{\Gamma}_{11}, \bar{\Gamma}_9 \bar{\Gamma}_{12})$ in $C_{3h}$, both the valence and the conduction bands lead to $\bar{\Gamma}_5 \bar{\Gamma}_6$ in $C_3$, which means that the band gap cannot close by changing a single parameter.
It seems to contradict the result in (c) in $C_{3h}$ that the $Z_2$ topological invariant changes at the band inversion in $C_{3h}$.
In this Appendix, we show that they are consistent and do not contradict.

To consider this problem, we consider an effective Hamiltonian for the case (c) $(\bar{\Gamma}_{9} \bar{\Gamma}_{12}, \bar{\Gamma}_{10} \bar{\Gamma}_{11})$ in $C_{3h}$.
We already have such a Hamiltonian in Eq.~\eqref{eq:hamiltonian_c3h_21}, up to the linear order in $\bm{k}$, but it is insufficient for the present purpose.
Therefore, we start from the following second-order effective Hamiltonian for the case (c) $(\bar{\Gamma}_{9} \bar{\Gamma}_{12}, \bar{\Gamma}_{10} \bar{\Gamma}_{11})$ satisfying the $C_{3h}$ symmetry:
\begin{equation}
    H_0(\bm{k}) = \mqty(
    \tilde{m} & 0 & \alpha + \beta & 0 \\
    0 & \tilde{m} & 0 & - \alpha^* + \beta^* \\
    \alpha^* + \beta^* & 0 & -\tilde{m} & 0\\
    0 & -\alpha + \beta & 0 & -\tilde{m} \\
    )
    \label{eq:hamiltonian_c3_c3h}
\end{equation}
with
\begin{align}
    \begin{split}
        \tilde{m} &= m + \frac{u_1 - u_2}{2} k^2,\\
        \alpha &= (v_1 + i v_2) (k_x - i k_y),\\
        \beta &= (u_3 + i u_4) (k_x + i k_y)^2,
    \end{split}
\end{align}
and $u_1, u_2, v_1, v_2, v_3, v_4$ are real parameters.
The pair of irreps of four bands at the TRIM is $(\bar{\Gamma}_9 \bar{\Gamma}_{12}, \bar{\Gamma}_{10} \bar{\Gamma}_{11})$, and the Hamiltonian preserves time-reversal symmetry:
\begin{equation}
    H_0(- \bm{k}) = \mqty(\sigma_y & \\ & \sigma_y) H_0^*(\bm{k}) \mqty(\sigma_y & \\ & \sigma_y).
\end{equation}
If we break $I C_6$ symmetry while preserving the $C_3$ symmetry in this situation, these two irreps become the same $\bar{\Gamma}_{5} \bar{\Gamma}_6$ and hybridize.
Namely, in the lowest order in $\bm{k}$, the following additional term is allowed by symmetry,
\begin{equation}
    \delta H = \mqty( & & & -A^* \\ & & A & \\ & A^* & & \\ - A & & & \\),
    \label{eq:symmetry_breaking_term}
\end{equation}
where $A$ is a complex constant.

\begin{figure}[t]
    \centering
    \includegraphics[width = 8cm]{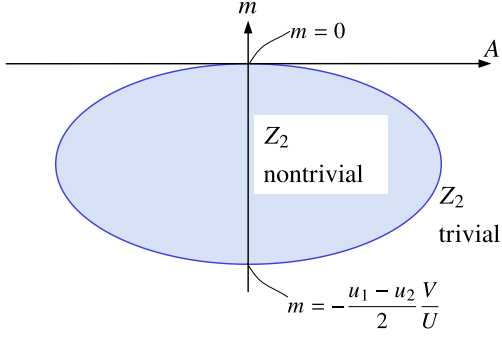}
    \caption{Phase diagram for the $C_{3h}$-symmetric model~\eqref{eq:hamiltonian_c3_c3h} with the symmetry-breaking term~\eqref{eq:symmetry_breaking_term} added. At $|A| = 0$, the model is $C_{3h}$-symmetric, and when $|A| \neq 0$, the symmetry is lowered to $C_3$. The phase diagram depends only on $|A|$. For illustration, $A$ is restricted to be real in the figure.}
    \label{fig:PD_C3h}
\end{figure}

We consider gap-closing of the $C_3$-symmetric Hamiltonian $H(\bm{k}) = H_0(\bm{k}) + \delta H$.
The eigenvalues have the form $\pm \sqrt{p \pm \sqrt{q}}$ by analytical calculations.
Here, $p$ and $q$ are functions of $\bm{k}$, but their expressions are lengthy, and we do not write them here.
Therefore, the gap closes when $p ^ 2 - q = 0$, which leads to
\begin{align}
      & p^2 - q = \det H \notag                                                                                        \\
    = & (|A|^2 + \tilde{m}^2 - |\alpha|^2 + |\beta|^2)^2 + 4 \tilde{m}^2 |\alpha|^2 + 4 [\Im(\alpha \beta^*)]^2 \notag \\
    = & 0.
\end{align}
Thus this equation requires
\begin{align}
    |A|^2 + \tilde{m}^2 - |\alpha|^2 + |\beta|^2 = 0, \label{eq:C_4} \\
    \tilde{m}^2 |\alpha|^2 = 0, \label{eq:C_5}                       \\
    \Im(\alpha \beta^*) = 0. \label{eq:C_6}
\end{align}
From Eq.~\eqref{eq:C_4} and Eq.~\eqref{eq:C_5}, we obtain
\begin{align}
    \tilde{m} & = 0                                        \\
    k^2       & = \frac{V \pm \sqrt{V^2 - 4 |A|^2 U}}{2 U}
    \label{eq:kk}
\end{align}
where $U = u_3^2 + u_4^2, V = v_1^2 + v_2^2$.
We solve these equations for $k$ and $m$ to clarify the conditions under which the band gap closes.

First, if $A = 0$, we get
\begin{equation}
    (k, m) = (0, 0), \left(\sqrt{\frac{V}{U}}, - \frac{u_1 - u_2}{2} \frac{V}{U} \right).
    \label{eq:k, m}
\end{equation}
In the first case, the gap closes at a single point with $k = 0$, while in the second case, considering Eq.~\eqref{eq:C_6}, the gap closes simultaneously at six points
\begin{align}
    \bm{k} & = \sqrt{\frac{V}{U}} (\cos \theta, \sin \theta),          \\
    \theta & = \frac{K}{3} + \frac{n \pi}{3} \quad (n \in \mathbb{Z}),
\end{align}
where $K = \arg [(v_1 + i v_2)(u_3 - i u_4)]$.
Second, if $A \neq 0$, the value of $|A|$ determines whether the gap closes.
When $|A|^2 < V / 4U$, by solving Eq.~\eqref{eq:kk}, we obtain two solutions for $\bm{k}: k = k^{(+)}, k^{(-)}$.
Therefore the gap can close at two values of $m: m = - \frac{u_1 - u_2}{2} {k^{(\pm)}}^2$, and at each value of $m$, the gap closes at six points in $k$-space.
When $|A|^2 > V / 4U$, the gap does not close.
Therefore, we can draw the phase diagram as shown in Fig.~\ref{fig:PD_C3h}.
In the model~\eqref{eq:hamiltonian_c3_c3h}, it is not determined which side of the phase diagram represents the topologically nontrivial phase.
For illustration, we set the interior of the phase boundary to be topologically nontrivial in Fig.~\ref{fig:PD_C3h}.

From this phase diagram, we see that the results for $C_{3h}$ and $C_3$ are entirely consistent.
In the $C_{3h}$-symmetric case ($A = 0$), the gap closing at $\bm{k} = 0, m = 0$ leads to a $Z_2$ topological phase transition.
Meanwhile, when $A$ becomes nonzero, the symmetry is lowered from $C_{3h}$ to $C_3$; then, the gap does not close at $\bm{k} = 0$, as expected from the same $C_3$ irreps $\bar{\Gamma}_5 \bar{\Gamma}_6$ at $\bm{k} = 0$.
Instead, the gap-closing points move away from $\bm{k} = 0$, and they are described by the $Z_2$ topological phase transition theory with no additional crystallographic symmetry in Refs.~\cite{Murakami2007-lt,Murakami2007-gu}.

\end{document}